\documentclass[fleqn,usenatbib]{mnras}

\usepackage{newtxtext,newtxmath}

\usepackage[T1]{fontenc}
\usepackage{ae,aecompl}

\usepackage{graphicx}	
\usepackage{amsmath}	
\usepackage{tablefootnote}

\defcitealias{2018A&A...615A..55T}{TM18}

\title[The effects of SFH in the SFR-$M_{*}$ relation of HII galaxies]{The effects of star formation history in the SFR-$M_{*}$ relation of HII galaxies}

\author[A. R. Lopes et al.]{
Amanda R. Lopes,$^{1}$\thanks{Contact e-mail: amandalopes1920@gmail.com}
Eduardo Telles$^{1}$
and Jorge Melnick$^{1,2}$
\\
$^{1}$Observat\'{o}rio Nacional, Rua Jos\'e Cristino, 77, Rio de Janeiro 20921-400, Brazil\\
$^{2}$European Southern Observatory, Av. Alonso de C\'{o}rdova 3107, Santiago, Chile
}

\date{Accepted XXX. Received YYY; in original form ZZZ}

\pubyear{2019}

\begin{document}
\label{firstpage}
\pagerange{\pageref{firstpage}--\pageref{lastpage}}
\maketitle

\begin{abstract}
We discuss the implications of assuming different star formation histories (SFH) in  the relation between star formation rate (SFR) and mass derived by the spectral energy distribution fitting (SED). Our analysis focuses on a sample of HII galaxies, dwarf starburst galaxies spectroscopically selected through their strong narrow emission lines in SDSS DR13 at $z<0.4$, cross-matched with photometric catalogs from GALEX, SDSS, UKIDSS and WISE. We modeled and fitted the SEDs with the code CIGALE adopting different descriptions of SFH. By adding information from different independent studies we find that HII galaxies are best described by episodic SFHs including an old (10 Gyr), an intermediate age (100$-$1000 Myr) and a recent population with ages < 10 Myr. HII galaxies agree with the SFR$-M_{*}$ relation from local star-forming galaxies, and only lie above such relation when the current SFR is adopted as opposed to the average over the entire SFH. The SFR$-M_{*}$ demonstrated not to be a good tool to provide additional information about the SFH of HII galaxies, as different SFH present a similar behavior with a spread of <0.1 dex.
\end{abstract}




\section{Introduction}
A widely used method to characterize galaxy star forming properties consists of fitting a model spectral energy distribution (SED) to observed photometry. In this technique assumptions must be made to create synthetic SEDs such as initial mass function, simple stellar population models, extinction law, star formation history and metallicity \citep[see][for reviews]{2011Ap&SS.331....1W, 2013ARA&A..51..393C}. Many authors attempted to verify the influence of different assumptions on this fitting process \citep[e.g.][]{2007ApJ...655...51W, 2009ApJ...699..486C, 2009ApJ...701.1839M, 2009ApJ...701.1765M, 2010MNRAS.407..830M, 2012MNRAS.422.3285P, 2013MNRAS.435...87M}. 

As an ingredient to create the model SEDs, the star formation history (SFH) can be described in simple parametric forms expressed by the relation between the star formation rate (SFR) and time. Many works applying simple analytical forms to model galaxies SFH can be found in the literature \citep[e.g.][]{2001ApJ...559..620P, 2013ApJ...770...64G, 2014arXiv1404.0402S, 2014A&A...561A..39B, 2014A&A...571A..72B, 2016ApJ...832....7A, 2017A&A...608A..41C, 2017A&A...605A..29S}. 

A common parameterized function is a single-component exponentially decreasing known as $\tau$-models. This class of parameterization suffers with the outshining effect for recent star formation in star-forming or/and high redshift galaxies, where the luminous young stars dominate the SED, often leading to an underestimation of old stars, thus stellar mass. Several works based on mock catalogs from semi-analytical or hydrodynamical galaxy formation models tried to improve the results in high-redshift galaxies proposing other SFH trends such as rising histories \citep[e.g.][]{2011MNRAS.410.1703F} and delayed models \citep[e.g.][]{2010ApJ...725.1644L}. Moreover, depending on the galaxy types different behaviors for the SFH were proposed. For example, \citet{2012A&A...541A..85M} studied the double exponential SFH in submillimetre galaxies, and \citet{2013MNRAS.431.2209B} used the exponential declining with additional burst to fit data from analogues of Lyman break galaxies. 

Similarly to the previous mentioned works we search for the best analytical SFH to a given type of galaxy. Our focus is on HII galaxies, compact dwarf starburst galaxies, characterized by a spectrum with weak, blue continuum and strong, narrow emission lines. In the local Universe they are considered the simplest starburst systems due to properties such as low mass, low oxygen abundance, low dust content and their low-density environments. Their low mass and metallicity make these objects local analogues of galaxies in the primordial Universe. Therefore, the study of their star formation activity can provide us with clues about the underlying physical processes in galaxies in earlier times.

HII galaxies are known for their recent star formation but different authors found observational \citep[e.g.][]{1997MNRAS.288...78T, 1997MNRAS.286..183T, 2004A&A...423..133W, 2011AJ....142..162L} and theoretical \citep[from simulations, e.g.][]{2004A&A...422...55P, 2016MNRAS.462.3739D} evidences to a multi-episode SFH, i.e. they all have an older stellar populations.
In this paper we elaborate on the work of \citet{2018A&A...615A..55T}. The authors assume three episodes of star formation for their multi-wavelength photometric data analysis. We use their selected galaxy sample to perform  SED-fitting tests assuming different descriptions of SFH. Due to limitations of the broad band SED fitting \citep[e.g.,][]{2012MNRAS.422.3285P,2014A&A...561A..39B,2015A&A...576A..10C,2016A&A...596A..11B,2019ApJ...873...44C}, we do not expect to find detailed, realistic SFH but to find the parametric SFH that best represents our observables. Our choice of SFHs is restricted to a few, and is constrained by other reasonable assumptions derived from the physical conditions from spectroscopic properties, such as low metallicity, gas consumption rate, and ionization parameter. These imply that HII galaxies have the presence of a significant amount of ionizing O and B stars ($M > 8M_{\odot}$), but cannot have kept the present rate of star formation through a Hubble time.  This allows a reasonable finite number of tests to be performed.

Based on the output from these tests we analyse, the SFR versus stellar mass ($M_{*}$), known as star-forming main sequence (MS) or SFR$-M_{*}$ relation. Many recent papers have studied the MS up to $z\sim$6 \citep[e.g.][]{2004MNRAS.351.1151B, 2007ApJ...660L..43N, 2007ApJS..173..267S, 2010ApJ...714.1740M, 2011A&A...533A.119E, 2012ApJ...754L..29W, 2015ApJS..219....8C, 2016ApJ...817..118T}. This correlation can be written as
\begin{equation}
\log(\mathrm{SFR}) = a \, \log(M_\mathrm{*})+b,
\label{eq.ms_def}
\end{equation}
where $a$ and $b$ are free parameters. \citet{2014ApJS..214...15S} compiled 25 papers from literature to discuss the evolution of both parameters as function of time. Other works suggest an extra dependence with mass, where low-mass galaxies should have a steeper slope than massive ones \citep[e.g.][]{2015ApJ...811L..12W, 2017A&A...597A..97M, 2019MNRAS.483.3213P}. 

Different authors \citep[e.g][]{2010ApJ...721..193P, 2015MNRAS.447.3548S, 2016MNRAS.457.2790T} suggest that the MS galaxies follow a single evolutionary track resulted from a combination of a myriad of physical processes such as infall of cold and hot gas, stellar mass loss, heating gas by stellar radiation, supernovae, active galactic nuclei. An alternative view has been presented by series of papers \citep{2013ApJ...770...64G, 2015ApJ...801L..12A, 2016ApJ...832....7A, 2016ApJ...833..251D, 2017ApJ...844...45O} where they argue that the MS is a natural outcome from most SFHs and has little astrophysical meaning. For our purpose we discuss the implications of different SFH in the SFR$-M_{*}$ relation for HII galaxies and compare it with the literature. Then we analyze if the MS can be used to infer any information about the SFH of HII galaxies.

The paper is divided in two complementary parts. The first part discusses the role of the SFH choice in the SED-fitting analysis, with a particular focus on seeking the best analytical SFH for HII galaxies. The second part takes advantage of the results from different SFHs to check if the SFR$-M_{*}$ relation can be used to retrieve any knowledge about the SFH.

The outline of the paper is as follows. \S 2 presents the data selection and its photometric properties. \S 3 describes the SED fitting procedure and \S 4 analyzes the results to search for the prefer SFH. \S 5 discusses how the SFH affects the derived SFR$-M_{*}$ relation for HII galaxies. Finally in \S 6 we summarize our conclusions. 

\section{Data description}
\label{sec:data_description}

HII galaxy sample was selected based on a cross-match between Sloan Digital Sky Survey (SDSS) Data Release 13 \citep{2017ApJS..233...25A} and the emission-LinePort table \citep[Portsmouth stellar kinematics and emission line flux measurements,][]{2013MNRAS.431.1383T}, according to the following criteria,
\begin{enumerate}
\item subclass classification of STARBURST, i.e. equivalent width of H$\alpha$ > 50\AA;
\item equivalent width of H$\beta$ > 30\AA;
\item line ratios that fall within the upper left panel of the star-forming regions in the BPT diagram \citep{1981PASP...93....5B, 2006MNRAS.372..961K}: $0<\log[\mathrm{OIII/H}\beta]<1.2$ and $-2.5<\log[\mathrm{NII/H}\alpha]<-0.8$.
\item redshifts in the range $0.005<z<0.4$ to avoid local giant HII regions in nearby galaxies. 
\end{enumerate}
Conditions (ii) and (iii) were imposed to ensure a sample of extreme star-forming galaxies with high excitation, low abundances, and low masses. Only objects with unique photometry in both far-ultraviolet (0.1528 $\mu$m) and near-ultraviolet (0.2271 $\mu$m) bands from Galaxy Evolution Explorer \citep[GALEX, e.g][]{2014Ap&SS.354..103B} are chosen, resulting in a final sample of 2728 galaxies. Detailed data description can be found in \citet[hereafter \citetalias{2018A&A...615A..55T}]{2018A&A...615A..55T}.

Besides the optical SDSS and ultraviolet regimes, our multi-wavelength photometry comprise bands in the near-infrared: Y (1.036 $\mu$m), J (1.250 $\mu$m), H (1.644 $\mu$m) and K (2.149 $\mu$m) from UKIRT Infrared Deep Sky Survey \citep[UKIDSS,][]{2007MNRAS.379.1599L} and the mid-infrared: W1 (3.4 $\mu$m) and W2 (4.6 $\mu$m) from Wide-Field Infrared Survey Explorer \citep[WISE,][]{2010AJ....140.1868W}.

In order to minimize systematic effects we used Petrosian magnitudes except for GALEX which we use model magnitudes. The aperture matching effects are minimum due to the compactness of our objects that have a median Petrosian radius in the SDSS r band of 2.8 arcsec (see Fig. 2 in \citetalias{2018A&A...615A..55T}).

\section{SED modeling and fitting}
We perform the SED modeling and fitting analysis with CIGALE\footnote{Code Investigating GALaxy Emission is available at https://cigale.lam.fr/ .} \citep[][]{2019A&A...622A.103B}. This is a package consisting of a series of modules that are combined to create a set of theoretical SEDs, then compared to the observed data \citep[e.g.][]{2009A&A...507.1793N, 2014ASPC..485..347R, 2014A&A...571A..72B, 2017A&A...608A..41C}. Each module corresponds to a physical component such as dust attenuation, SFH and single stellar population model. These components have their own parameters to be chosen. The comparison between theoretical and observed data results in a best-fit model and a Bayesian-like analysis, obtained by the probability distribution function of each parameter. 

\begin{figure}
\centering
\includegraphics[width=0.47\textwidth]{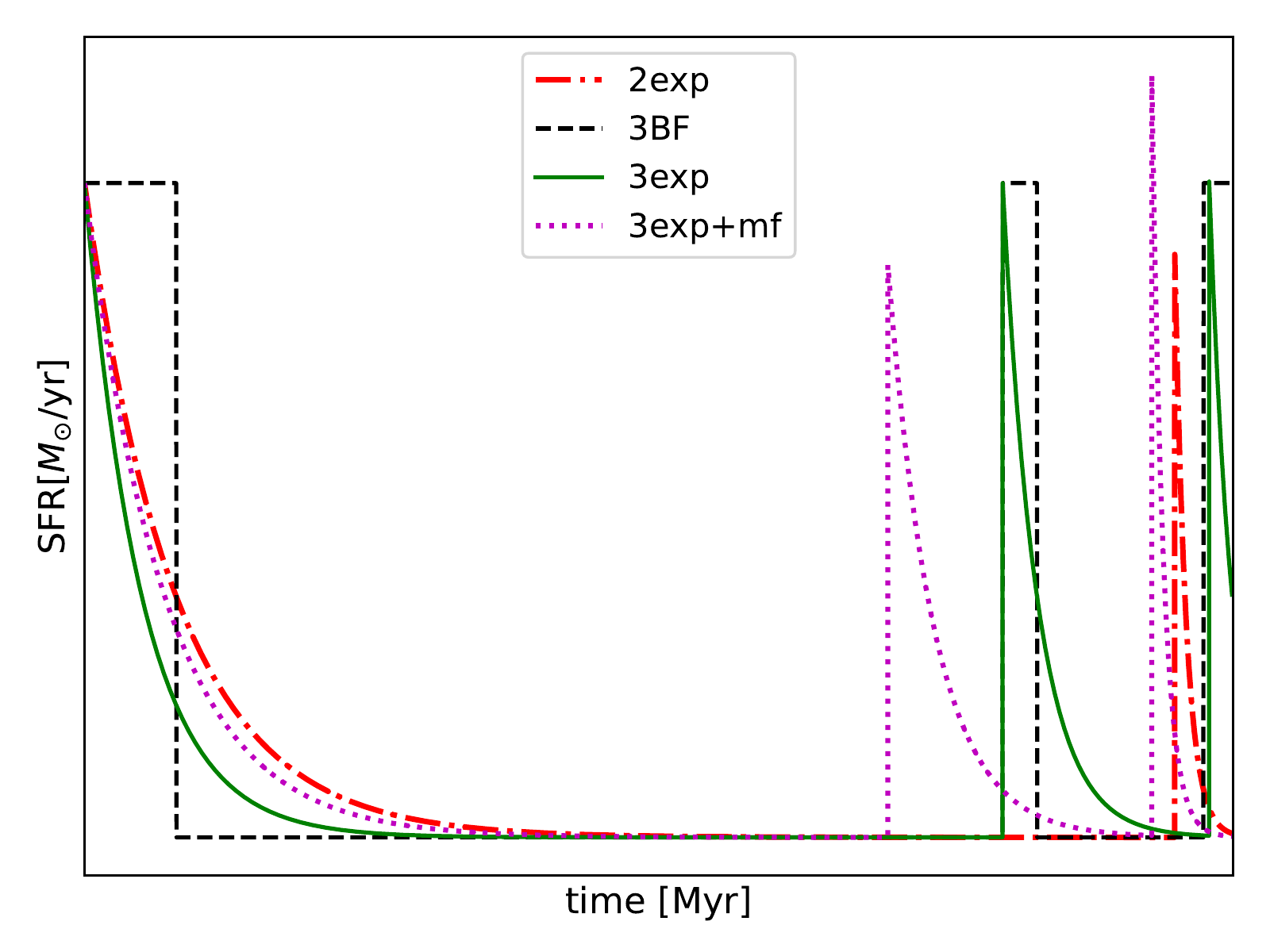}
\caption{Illustration of the SFH used in our tests: double (red dash-dotted line), triple exponentially declining (green solid line), 3BF (black dashed line) and triple exponential with varying amplitude amongst the episodes of star formation (magenta dotted line).}
\label{fig:sfh_examples}
\end{figure}

A key factor for our purpose and one of the main reasons for choosing CIGALE is its modular structure that facilitates the introduction of new SFH modules. 
Since our goal is to test SFHs, we set all other parameters with the same input values as described in Table \ref{tab:cigale_cte}. The behavior of the SFHs applied in our work is shown in Fig.~\ref{fig:sfh_examples}. They have different numbers of star formation episodes and/or shapes as we discuss next. 
Throughout this paper, we will adopt the flat $\Lambda$CDM cosmology with $H_{0}$= 71 km/s/Mpc.

\begin{table}
\caption{List of CIGALE input parameters unchanged in all tests. For more details about the choice of these values, see \citetalias{2018A&A...615A..55T}.}
\centering
\label{tab:cigale_cte}
\begin{tabular}{lc}
\hline
Parameter                 & Value \\
\hline
\multicolumn{2}{c}{Module: Simple Stellar Population (SSP)} \\
\hline
Model         & \citet{2003MNRAS.344.1000B}\\
Initial Mass Function (IMF)           & \citet{2003PASP..115..763C} \\ 
Metallicity   &  0.008 \\
Separation between & \\
young and old pop [Myr] & 10 \\
\hline
\multicolumn{2}{c}{Module: Nebular Emission} \\
\hline
Ionization parameter & $\log U=-2.0$\\
LyC photons escaping the galaxy & 0.0 \\
LyC photons absorbed by dust & 0.0 \\
Line width in km/s & 300.0 \\
\hline
\multicolumn{2}{c}{Module: Dust Attenuation} \\
\hline
Name & \citet{2000ApJ...533..682C}\\
E(B$-$V)$_{\mathrm{line}}$\tablefootnote{E(B$-$V)$_\mathrm{line}$ represents the colour excess of the nebular lines light for both the young and old population} & 0., 0.05, 0.1, 0.15, 0.2,\\
&  0.25, 0.3, 0.35, 0.4, 0.45\\ 
E(B$-$V)$_\mathrm{factor}$\tablefootnote{E(B$-$V)$_\mathrm{factor}$ is the reduction factor to be applied on E(B$-$V)$_\mathrm{line}$ to compute E(B-V)s from the stellar continuum attenuation.} & 0.44, 1 \\
Amplitude of the UV bump & 1.0 \\
Slope delta of the power law & -0.5\\
\hline
\multicolumn{2}{c}{Module: Dust Emission}\\
\hline
Dust template & Updated models of\\
 & \citet{2007ApJ...657..810D}\\
Mass fraction of PAH & 0.47, 1.12\\
Minimum radiation field & 0.1 \\
IR power law slope & 2.0\\
\hline
AGN template & NONE\\
\hline
Radio & NONE\\
\hline
\end{tabular}
\end{table}

\subsection{Star formation histories}
\label{subs:sfh_def}
Following papers on spectral and imaging analysis of HII galaxies such as \citet{1997MNRAS.288...78T}, \citet{2004A&A...423..133W} and \citet{2011AJ....142..162L} that found evidence of underlying older stellar populations, \citetalias{2018A&A...615A..55T} proposed a SFH composed of three main episodes of star formation, given by
\begin{equation}
\mathrm{SFR}(t) =
\begin{cases}
C \quad \mathrm{for} \quad t_{\mathrm{old}}-\tau_{\mathrm{old}}<t<t_{\mathrm{old}} \\
C \quad \mathrm{for} \quad t_{\mathrm{int}}-\tau_{\mathrm{int}}<t<t_{\mathrm{int}}\\
C \quad \mathrm{for} \quad t_{\mathrm{y}}-\tau_\mathrm{y}<t<t_{\mathrm{y}}\\
0 \quad \mathrm{otherwise},
\end{cases}
\label{eq:melnick}
\end{equation}
where $C$ is a constant value estimated by the code, $t_{\mathrm{i}}$ and $\tau_{\mathrm{i}}$ are the starting time and duration of each star formation event. The subscript ($i=$ old, int, y) indicates an old, an intermediate age and a young (ionizing) stellar population. This expression leads to the amount of stars created depending only on $\tau_{i}$ interval. This is not a standard SFH module in CIGALE and it was created by M. Boquien for that paper.  

Throughout this paper we will refer to this SFH as three burst function (hereafter 3BF). We will also maintain the labels, old, intermediate and young, to describe stellar populations with ages $\sim$10 Gyr, $\sim$100$-$1000 Myr and $\lesssim$10 Myr, respectively. 

Alternatively we can assume two main episodes of star formation with the first one being very extended. By imposing no intermediate age stellar population in the 3BF description, we have a two burst function, hereafter 2BF.

Another simple way of modeling a SFH with two episodes of star formation is assuming two decaying exponentials combined as \citep[e.g.][]{2019A&A...622A.103B},
\begin{equation}
\mathrm{SFR}(t) \propto 
\begin{cases} 
\exp(-t/\tau_{0}) \quad \mathrm{for} \quad t < t_{0}-t_{1} \\
\exp(-t/\tau_{0})+ k \exp(-t/\tau_{1})\quad \mathrm{for} \quad t \geq t_{0}-t_{1},
\end{cases}
\label{eq:double.exp}
\end{equation}
where $t_{1}$ is the age of the second episode of star  formation relative to $t_{0}$, $\tau_{0}$ and $\tau_{1}$ are the e-folding times of the old and recent stellar populations, and $k$ represents the relative amplitude of the second exponential defined by
\begin{equation}
k = \left(\frac{f_{\mathrm{y}}}{ 1-f_{\mathrm{y}}}\right) \frac{\sum_{t_{i}=0}^{t_{i} =t_{1}-1} \exp(-t_{i}/\tau_{1})}{\sum_{t_{i}=t_{1}-t_{2}-1}^{t_{i}=t_{1}-1} \exp(-t_{i}/\tau_{2})},
\label{eq:double.frac}
\end{equation}
with $f_{\mathrm{y}}$ being the mass fraction of the late stellar population. We will label this description as 2exp.

Finally, we considered two SFHs that expand upon the 3BF. First, we altered the description assuming three exponential declining functions instead of constant SFR described by,
\begin{equation}
\mathrm{SFR}(t) \propto
\begin{cases} 
\exp(-t/\tau_{\mathrm{old}}) \quad \mathrm{for} \quad t < t_{\mathrm{old}}\\
\exp(-t/\tau_{\mathrm{int}}) \quad \mathrm{for} \quad  t < t_{\mathrm{int}}\\
\exp(-t/\tau_{\mathrm{y}}) \quad \mathrm{for} \quad t < t_\mathrm{y}, \\
\end{cases} 
\label{eq:triple.exp}
\end{equation}
where $t_{\mathrm{old}}$, $t_{\mathrm{int}}$ and $t_{\mathrm{y}}$ are the ages of the oldest star in each stellar population, old, intermediate and young, respectively, with their associated interval of star formation, $\tau_{\mathrm{old}}$, $\tau_{\mathrm{int}}$ and $\tau_{\mathrm{y}}$. By definition this function imposes that the maximum SFR in every star formation event to be the same. We will refer to this SFH as 3exp.

Secondly, we introduce a triple exponential with varying amplitude given by
\begin{equation}
\mathrm{SFR}(t) \propto 
\begin{cases} 
\exp(-t/\tau_{\mathrm{old}})\quad \mathrm{for} \quad t < t_{\mathrm{old}}\\
k_{1}\exp(-t/\tau_{\mathrm{int}}) \quad \mathrm{for} \quad t < t_\mathrm{int} \\
k_{2}\exp(-t/\tau_{\mathrm{y}}) \quad \mathrm{for} \quad t < t_\mathrm{y}, \\
\end{cases} 
\label{eq:triple.exp.mf}
\end{equation}
where $k_{1}$ and $k_{2}$ are the amplitudes of intermediate and young age population defined by
\begin{equation}
k_{1} = \left(\frac{f_{\mathrm{int}}}{1-f_{\mathrm{int}}-f_{\mathrm{y}}}\right)\frac{\sum_{t_{i}=0}^{t_{i}=t_{\mathrm{old}}-1} \exp(-t_{i}/\tau_{\mathrm{old}})}{\sum_{t_{i}=t_\mathrm{old}-t_\mathrm{int}-1}^{t_{i}=t_{\mathrm{old}}-1} \exp(-t_{i}/\tau_{\mathrm{int}})},
\label{eq:k1.def}
\end{equation}
and
\begin{equation}
k_{2} = \left(\frac{f_{\mathrm{y}}}{f_{\mathrm{int}}}\right)\frac{\sum_{t_{i}=t_\mathrm{old}-t_\mathrm{int}-1}^{t_{i}=t_{\mathrm{old}}-1} k_{1}\exp(-t_{i}/\tau_{\mathrm{int}})}{\sum_{t_{i}=t_\mathrm{old}-t_\mathrm{y}-1}^{t_{i}=t_{\mathrm{old}}-1} \exp(-t_{i}/\tau_{\mathrm{y}})},
\label{eq:k2.def}
\end{equation}
with $f_{\mathrm{int}}$ and $f_{\mathrm{y}}$ being the mass fractions of the intermediate and young populations, respectively. As this SFH follows a 3exp behavior added by two new free parameters in the form of mass fraction, we will label it 3exp+mf. 

\section{The Star Formation History of HII galaxies}
\label{sec.sfh}

We need to combine observational knowledge from other independent studies to assist in the search for the best analytical SFH. For example, detailed studies of nearby objects reveal that regardless of morphological type, all dwarf galaxies contain old (10$-$12 Gyr) stars \citep[e.g.,][]{2004ApJ...610L..89G,2008AJ....135.1106G,2010ApJ...708L.121D,2012AIPC.1480..172G}. Therefore, independently of the shape of the SFH, the star formation should start around 10$-$12 Gyr.

HII galaxies are known for their strong current star formation, dominated by a stellar population of <10 Myr that must be included in the description of these galaxies.

The constraints for the SFH are:
\begin{enumerate}
    \item starts at $t_\mathrm{old}=10$ Gyr;
    \item has young, ionizing, massive stellar population at $t_\mathrm{y}<10$ Myr.
    \item requires an intermediate age population at $100<t_\mathrm{int}/\mathrm{Myr}<1000$. 
\end{enumerate}

\citetalias{2018A&A...615A..55T} showed that models assuming a 3BF SFH description with their set of input parameters provides a good fit to the photometric data. For this reason we apply this function with a similar range of input values, and use it as a base of comparison for other tests. This run is labeled as \texttt{test0}. 

\begin{table*}
\caption{Input parameters for the SFH modules in each test with their respective total number of SED models.}
\centering
\label{tab:sfh_input}
\begin{tabular}{ccl|c}
\hline\hline
Tests  & SFH & Input parameters & Number of \\
       &     &                  & SED models \\
\hline
\texttt{test0} & 3BF & $t_\mathrm{old}$=10 Gyr ; $\tau_\mathrm{old}$=[100,300,500] Myr & 133200 \\
               &     & $t_\mathrm{int}$=[100,500,1000] Myr ; $\tau_\mathrm{int}$=[10,50,100,200,300] Myr & \\
               &     & $t_\mathrm{y}$=[5,10] Myr ; $\tau_\mathrm{y}$=5 Myr & \\
\hline               
\texttt{test1} & 2BF & $t_\mathrm{old}$=10 Gyr ; $\tau_\mathrm{old}$=[100,300,500,1000,2000,3000,4000,5000,6000,7000,8000,9000,9500] Myr & 38480 \\ 
               &     & $t_\mathrm{int}$=0 ; $\tau_\mathrm{int}$=0 & \\ 
               &     & $t_\mathrm{y}$=[5,10] Myr ; $\tau_\mathrm{y}$= 5 Myr & \\ 
\hline
\texttt{test2} & 2exp & $t_\mathrm{0}$=10 Gyr ; $\tau_\mathrm{0}$=[100,300,500,1000,2000,3000,4000,5000,6000,7000,8000,9000,9500] Myr & 115440 \\
& & $t_\mathrm{1}$=[5,10] Myr ; $\tau_\mathrm{1}$= 5 Myr &\\
& & $f_\mathrm{y}$=[0.1,0.5,0.9] & \\
\hline
\texttt{test3} & 3exp & $t_\mathrm{old}$=10 Gyr ; $\tau_\mathrm{old}$=[100,300,500] Myr & 133200 \\ 
               &     & $t_\mathrm{int}$=[100,500,1000] Myr ; $\tau_\mathrm{int}$=[10,50,100,200,300] Myr & \\
               &     & $t_\mathrm{y}$=[5,10] Myr ; $\tau_\mathrm{y}$= 5 Myr & \\
\hline
\texttt{test4} & 3exp+mf & $t_\mathrm{old}$=10 Gyr ; $\tau_\mathrm{old}$=[100,300,500] Myr & 5594400 \\ 
               &     & $t_\mathrm{int}$=[100,500,1000] Myr ; $\tau_\mathrm{int}$=[10,50,100,200,300] Myr & \\
               &     & $t_\mathrm{y}$=[5,10] Myr ; $\tau_\mathrm{y}$= 5 Myr & \\
               &    & $f_\mathrm{int}$=[0.01,0.05,0.1,0.2,0.3,0.4,0.5] ; $f_\mathrm{y}$=[0.01,0.05,0.1,0.2,0.3,0.4] & \\
\hline\hline
\end{tabular}
\end{table*}

Table \ref{tab:sfh_input} summarizes the parametrization of our test runs with CIGALE. Our aim is to test the following effects:
\begin{itemize}
    \item number of episodes of star formation, e.g. \texttt{test1} and \texttt{test2} assume one extended event of star formation followed by a recent one, instead of 3 events (\texttt{test0});
    \item shape of each episode, e.g. \texttt{test3} assumes an exponential decay behavior rather than a constant one; 
    \item \texttt{test4} allows different intensities in the SFR for each stellar population.
\end{itemize}
We tested several other ranges of input parameters, but the results were consistent with the ones presented in this paper. 

\begin{figure}
\centering
\includegraphics[width=0.47\textwidth]{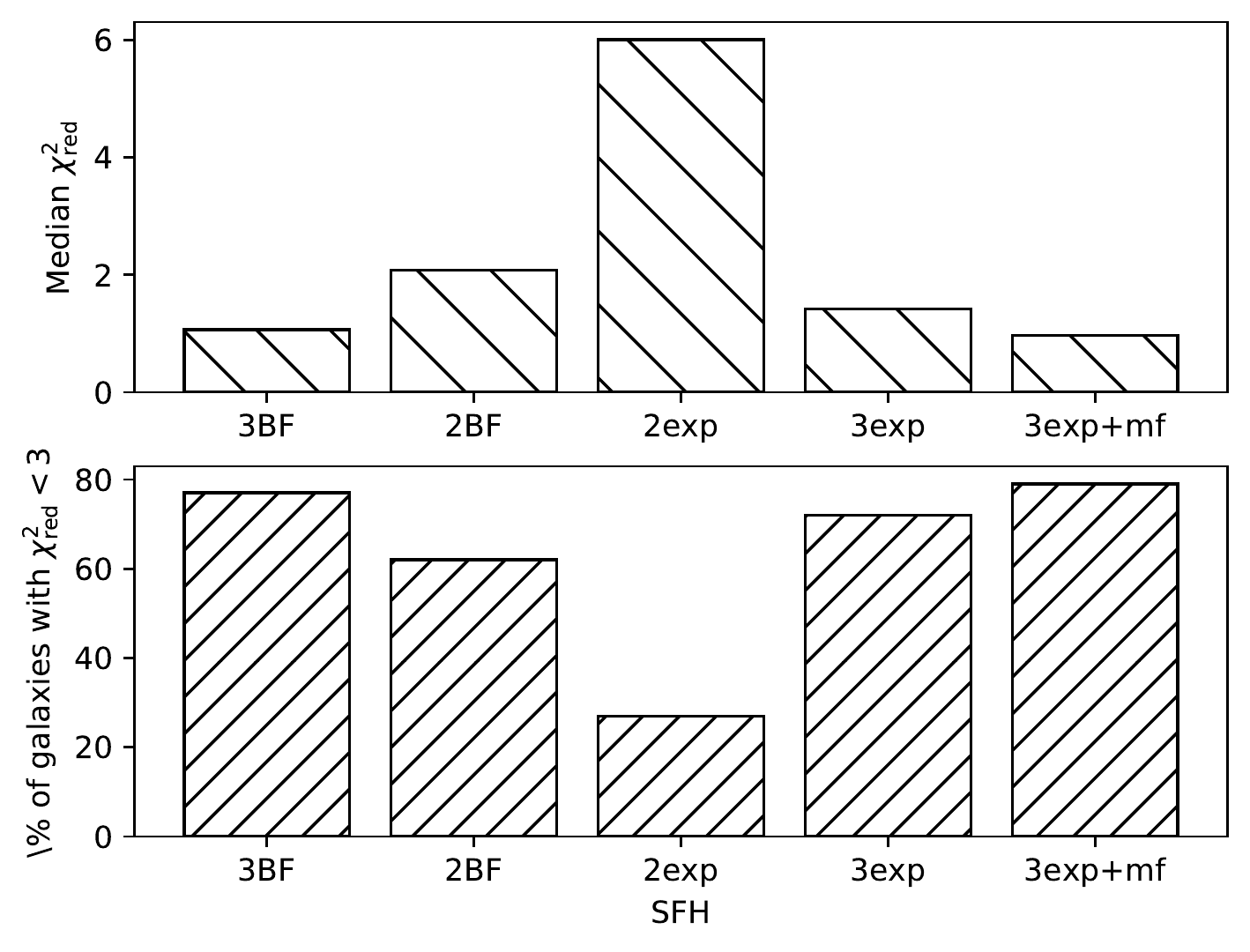}
\caption{Median $\chi^{2}_\mathrm{red}$ (top panel) and the percentage of galaxies with $\chi^{2}_\mathrm{red}<3$ (bottom panel) for our set of tests.}
\label{fig:stats_tests}
\end{figure}

The goodness of the fit is assessed by the $\chi^{2}_\mathrm{red}$. Figure \ref{fig:stats_tests} presents an statistical overview of the results from all test. Both panels in this figure show that tests assuming three main episodes of star formation have a similar performance with median $\chi^{2}_\mathrm{red}\sim 1$ and $\approx80$\% of the galaxies with a good fit, whereas SFHs with two episodes of star formation have worse performance. We restrict our analysis in the next subsections to galaxies with $\chi^{2}_\mathrm{red}$ <3 for all tests.

The worst results are derived by 2exp SFH, which fits only $\sim30\%$ of galaxies with $\chi^{2}_\mathrm{red}< 3$. This is significantly worse than the output based on 2BF function that finds twice the number of galaxies with $\chi^{2}_\mathrm{red}< 3$. However, the differences between both tests are mainly in the shapes of the SFHs. Similarly, 3BF provides better results than 3exp, but with a lesser difference in the efficiency.

\begin{figure}
\centering
\includegraphics[width=0.47\textwidth]{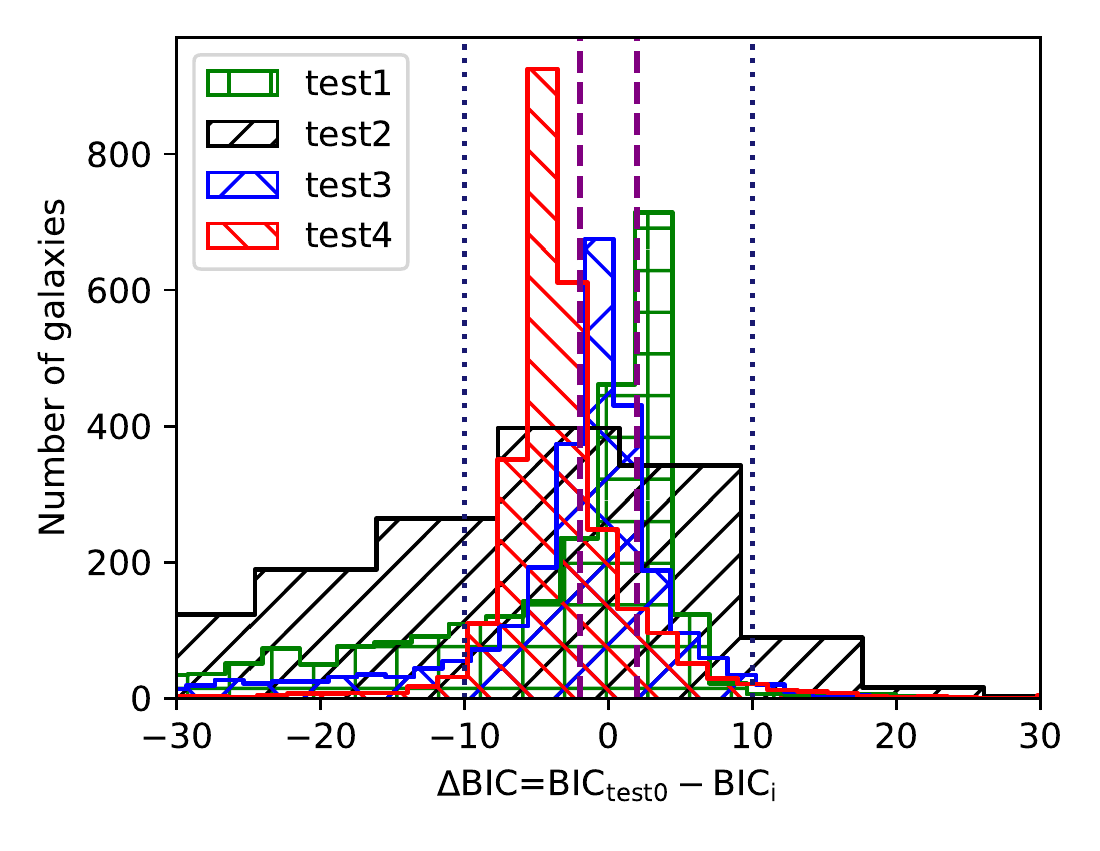}
\caption{Histogram of BIC for the different tests (1 to 4) in comparison to \texttt{test0}. For |$\Delta$BIC|<2, the distinct SFHs present a similar efficiency to fit the data (delimited by purple dashed line). For a more conservative limit (delimited by dark blue dotted line), we set |$\Delta$BIC|<10. The higher the $\Delta$BIC value the bigger the difference in the efficiency between the compared models, with the corresponding sign informing which of the models is more efficient. A positive sign means that \texttt{test}i has a better performance than \texttt{test0}, while a negative sign has the opposite meaning. For example \texttt{test2} has worse efficiency than \texttt{test0} because $\Delta$BIC has a broad distribution towards the negative side of the plot.}
\label{fig:bic}
\end{figure}

An alternative statistical approach to discuss the goodness of the fit is the Bayesian Information Criterion \citep[BIC, e.g.][]{2007MNRAS.377L..74L} that penalizes heavily the model with more parameters. The BIC checks the improvement (or lack of it) in the fitting analysis by adding or subtracting parameters in our SFH. It can be estimated using the following expression
\begin{equation}
\mathrm{BIC} = \chi^{2}+k\ln{N},
\label{eq.BIC}
\end{equation}
where $\chi^{2}$ is the non-reduced $\chi^{2}$ of the fit, $k$ is the number of free parameters to create the models, and $N$ is the number of bands fitted that varies from 7 to 13. The only difference between the tests is the assumed SFH with its number of parameters that range from 2 to 6. We use the relative difference between the BIC from \texttt{test0} and the other tests represented as
\begin{equation}
\Delta\mathrm{BIC} = \mathrm{BIC}_\mathrm{test0} - \mathrm{BIC}_\mathrm{testi},
\label{eq.delta_bic}
\end{equation}
with \texttt{test}i = \{\texttt{test1}, \texttt{test2}, \texttt{test3} or \texttt{test4}\}. 
This methodology implies that models with |$\Delta$BIC|<2 have similar fitting performance, i.e. the different SFH prescriptions and number of parameters do not greatly affect the fit. The influence is considered moderate if 2<|$\Delta$BIC|<10, while |$\Delta$BIC|>10 means a strong influence from SFH expression. The sign of $\Delta$BIC informs us which of the compared models is more efficient. A negative sign means that \texttt{test0} has a better performance than \texttt{test}i, while a positive one means \texttt{test}i provides a better fit.

Figure~\ref{fig:bic} shows histograms of $\Delta$BIC from tests 1 to 4 in respect to \texttt{test0}. There are important features to notice in this figure: where the peak of distribution is located, how broad $\Delta$BIC is, and if the sign is positive or negative. 

Three of the tests have similar behaviors presenting a tall and thin distribution around a peak in the regions between a weak to moderate dependence on the SFH applied. The main difference is the location of the peak: \texttt{test1} is $\sim 3$, \texttt{test3} is $\sim 0$ and \texttt{test4} is $\sim -2$. Therefore, \texttt{test3} is as good as \texttt{test0}, while the increase of 2 parameters in \texttt{test4} does not significantly improve the fit. \texttt{Test1} has a slightly advantage in respect to \texttt{test0}, due to fitting only 2 parameters, but it is not strong enough for us to eliminate the intermediate age stellar population from the SED modeling.

The one exception is \texttt{test2}, in which the $\Delta$BIC distribution is broader, extending to values lower than -20, with a peak not as well defined as the other tests. The negative sign means that \texttt{test0} provides a better fit solution, agreeing with the $\chi^2_\mathrm{red}$ analysis (Figure \ref{fig:stats_tests}).

Combining $\chi^{2}_\mathrm{red}$ and BIC analysis it would be straightforward to conclude that HII galaxies need at least three main stellar populations to be well modeled. Also, \texttt{test0}, \texttt{test3} and \texttt{test4} seem provide a similar good fit. However, we should take these results with caution. First, we need to deepen our analysis by discussing the SED procedure limitations.

\begin{figure*}
    \centering
    \includegraphics[width=\textwidth]{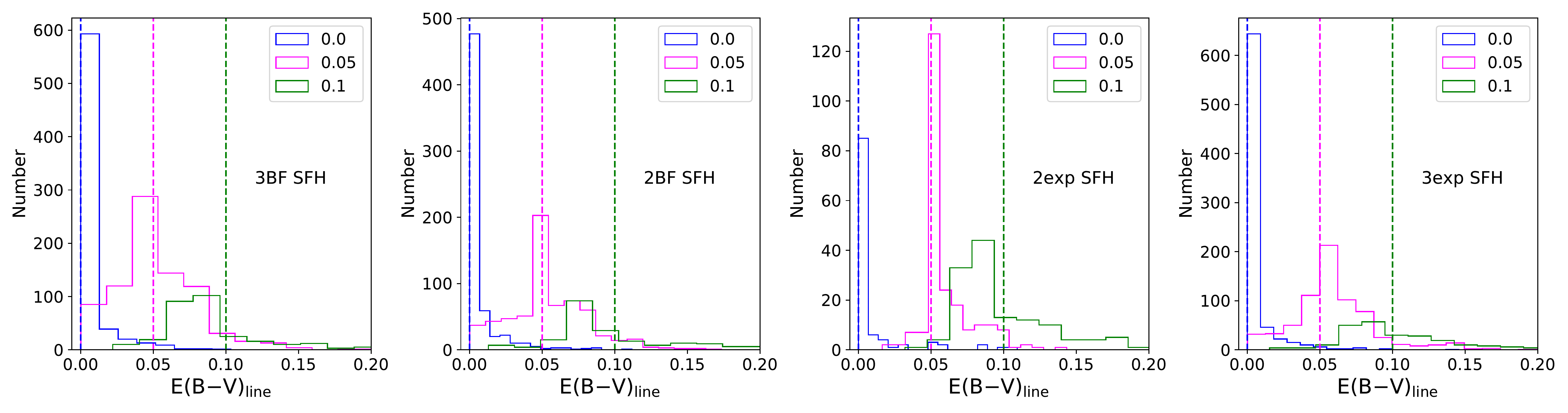}
    \caption{Histograms of the estimated E(B$-$V) using mock catalogues for different SFHs (panels from left to right: 3BF, 2BF, 2exp and 3exp). Each color represent a true value and its respective output distribution. To guide the eyes, we illustrate the position of each true value by dashed lines. Results from 3exp+mf SFH were not included but they are very similar to the ones from 3exp.}
    \label{fig:ebv}
\end{figure*}

\begin{figure*}
    \centering
    \includegraphics[width=\textwidth]{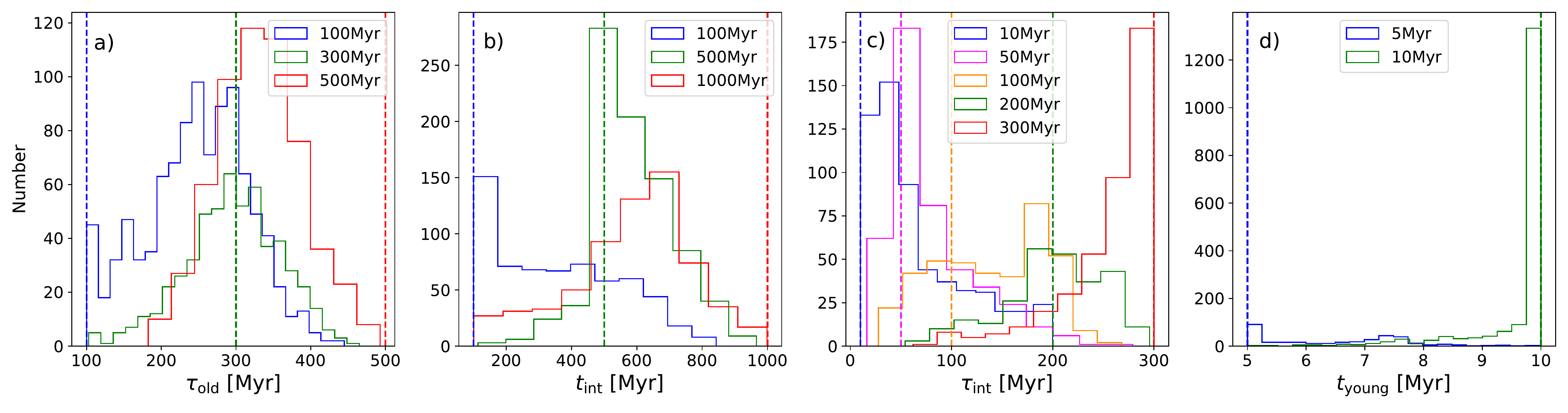}
    \caption{Histograms of the estimated parameters of 3BF SFH using mock catalogues. Parameters $t_\mathrm{old}$ and $\tau_\mathrm{young}$ are fixed. Each color represent a true value and its respective output distribution. To guide the eyes, we illustrate the position of each true value by dashed lines.}
    \label{fig:mock_3bf}
\end{figure*}

\subsection{Mock catalogue analysis}
\label{subsec.mock_analysis}
In order to verify the reliability of the derived physical properties, we use a mock catalogue to test our results \citep[see][for a more detailed description, and  similar analysis with mock catalogues in CIGALE]{2011A&A...525A.150G,2012A&A...539A.145B}. This artificial catalogue is created based on the best fit for each object. The best fit of each physical parameter is set as a ``true" measurement. Then to simulate new observations, it is added noise into the fluxes of this new catalogue. Finally, these synthetic observations are analyzed following the same procedure as the original data. The inferred physical properties are compared to their ``true" values. This process is done automatically in CIGALE by setting \texttt{mock\_flag=True} in the configuration file.

The degeneracy between age, metallicity and  extinction can be a source of uncertainties in the SED fitting procedure. However, the metallicity is set as a fixed value for all galaxies, estimated by spectroscopic data. All tests have $\sim85\%$ of the sample best-fitted by E(B-V)$\lesssim0.1$, therefore the mock analysis in this range of values should be representative of the whole sample. The peak of the estimated E(B-V) distribution has a good consistency with the expected values of 0 and 0.05 in all four panels in Figure \ref{fig:ebv}. The worse case is for inputs of 0.1, which the peak of outputs is offset to 0.07, but these values are so close to each other that in practice there is no difference. The results from 3exp+mf SFH were not included because they are similar to 3exp SFH. The small derived E(B-V)s, $\lesssim 0.1$, agrees with our expectations, as HII galaxies are known to have low dust content. The good reliability of E(B-V) for all tests indicates that most uncertainties must come from the parameters in the SFH.

Before individual discussions on the SFH parameters from each test, we need to make important remarks about known limitations of broadband SED-fitting technique and our choices of input parameters. For SFHs described by exponential functions, we tested  $\tau_\mathrm{y}$ ranging from 5 Myr to 15 Gyr, but no improvement was seen. In these extra tests, the best-fit $\tau_\mathrm{y}$ was $\sim 5$ Myr for 90\% of our galaxy sample. As the results were mostly unchanged, we opted to use simplest combination of input parameters with $t_\mathrm{y}$= [5, 10] Myr, and $\tau_\mathrm{y}=5$ Myr.

\begin{figure*}
    \centering
    \includegraphics[width=\textwidth]{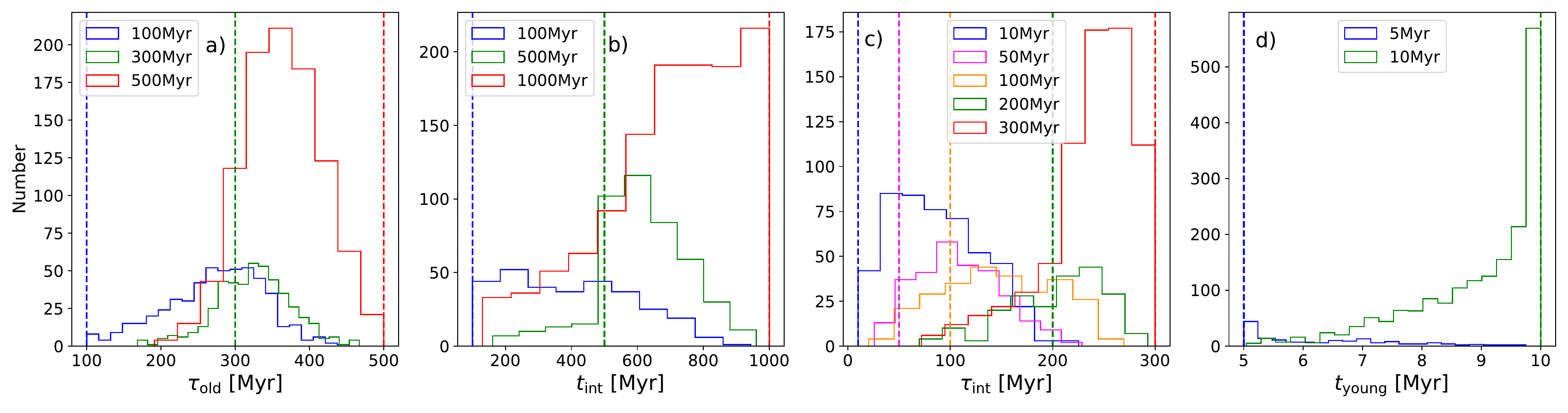}
    \caption{Histograms of the estimated parameters of 3exp SFH using mock catalogues. Parameters $t_\mathrm{old}$ and $\tau_\mathrm{young}$ are fixed. Each color represent a true value and its respective output distribution. To guide the eyes, we illustrate the position of each true value by dashed lines.}
    \label{fig:mock_3exp}
\end{figure*}
\begin{figure*}
    \centering
    \includegraphics[width=0.6\textwidth]{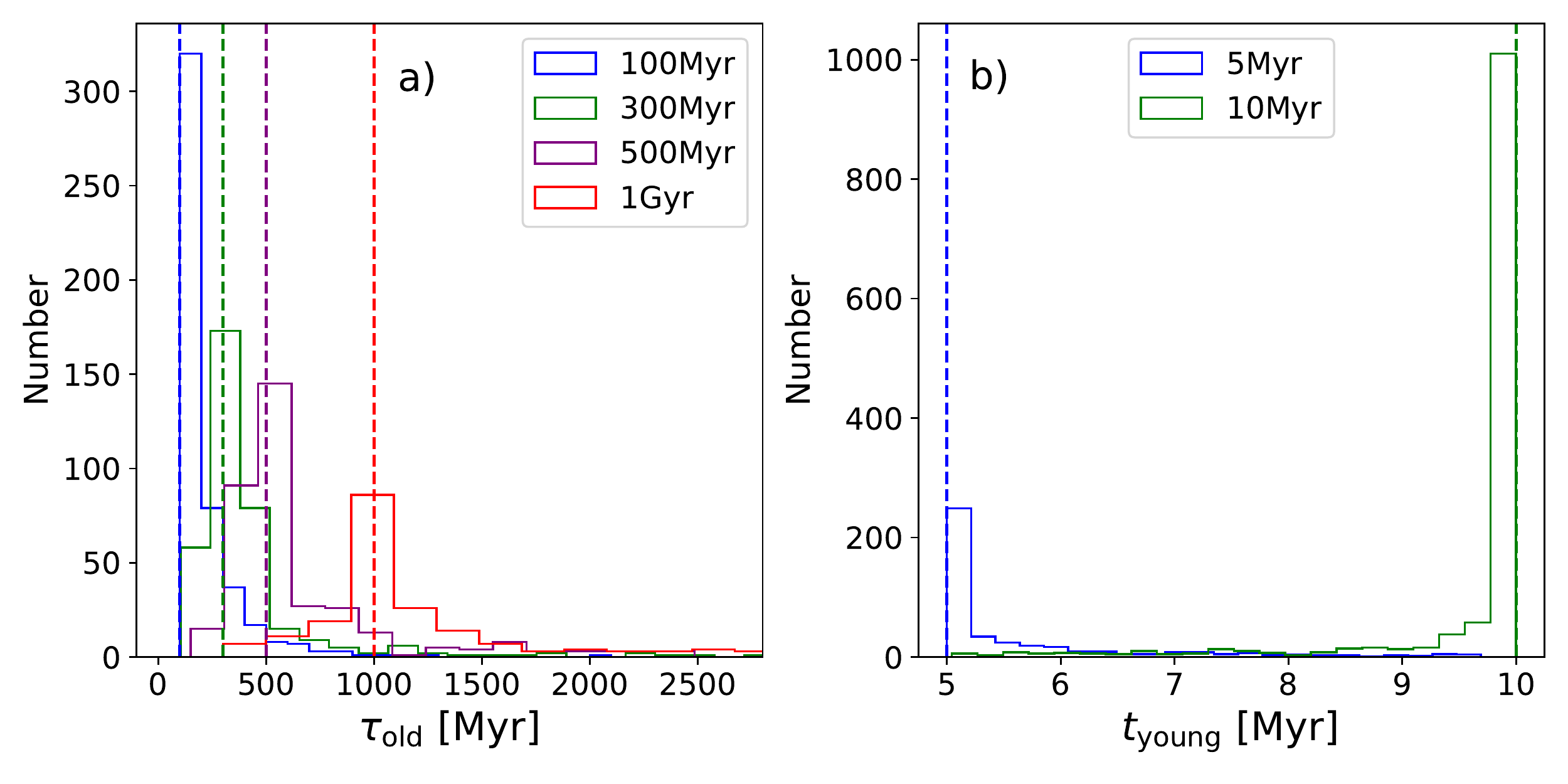}
    \caption{Histograms of the estimated parameters of 2BF SFH using mock catalogues. As 82\% of the sample is best-fitted by $\tau_\mathrm{old}$ up to 1 Gyr, we chose to exclude other values from the plot. Parameters $t_\mathrm{old}$ and $\tau_\mathrm{young}$ are fixed. Each color represent a true value and its respective output distribution. To guide the eyes, we illustrate the position of each true value by dashed lines.}
    \label{fig:mock_2bf}
\end{figure*}

Figure \ref{fig:mock_3bf} presents the results of the parameters from 3BF SFH using mock catalogues. In these figures, and the ones that follow, the vertical lines correspond to the input ("true") values and the histograms are the output from the mock run. The better agreement of the histogram peaks with the vertical lines, the better is the retrieval of the input values, and the more robust are the results. For the old stellar population with age fixed at 10 Gyr, Figure \ref{fig:mock_3bf}a shows that a star formation duration of 300 Myr is better reproduced than 100 Myr or 500 Myr. Figure \ref{fig:mock_3bf}b demonstrates that stellar populations of 100 Myr and 500 Myr have robust measurements, whereas populations of 1 Gyr have underestimated ages. In Figure \ref{fig:mock_3bf}c, we see that the duration of formation of these intermediate age stars is fairly well reproduced, but with a overlap distribution between 10 Myr and 50 Myr, and 100 Myr and 200 Myr. Finally, Figure \ref{fig:mock_3bf}d exhibits a good agreement between estimate and true ages of stars with 10 Myr.

Figure \ref{fig:mock_3exp} shows the reproduction of the input values from 3exp SFH. In Figure \ref{fig:mock_3exp}a, the duration of the oldest episode of star formation has better results for 300 Myr, similar to 3BF SFH. Figure \ref{fig:mock_3exp}b shows good estimates of ages for stars with 500 Myr and 1 Gyr. In Figure \ref{fig:mock_3exp}c, we find no good agreement for the intervals of intermediate age star formation, with most intervals being overestimated, except for 300 Myr where is underestimate. Figure \ref{fig:mock_3exp}d reveals that stars with 10 Myr have slightly worse reproduction of their ages for 3exp in comparison to 3BF SFH, shown by a broader range outputs.

Figure \ref{fig:mock_2bf} presents the parameter results from 2BF SFH. By excluding the intermediate age stellar population, we have only 2 parameters to fit and analyze in 2BF, the duration of the formation of old stars $\tau_\mathrm{old}$ and the age of young population $t_\mathrm{young}$. In Figure \ref{fig:mock_2bf}a, we select $\tau_\mathrm{old}$ inputs from 100 Myr to 1 Gyr, as these values are the best-fit of 82\% of galaxies, therefore representative of the entire sample. This panel illustrates how well the peak of derive quantities (histograms) reconciles with each true value (dashed lines). Figure \ref{fig:mock_2bf}b demonstrates that both input ages are well reproduced by SED-fitting procedure.

\begin{figure*}
    \centering
    \includegraphics[width=0.9\textwidth]{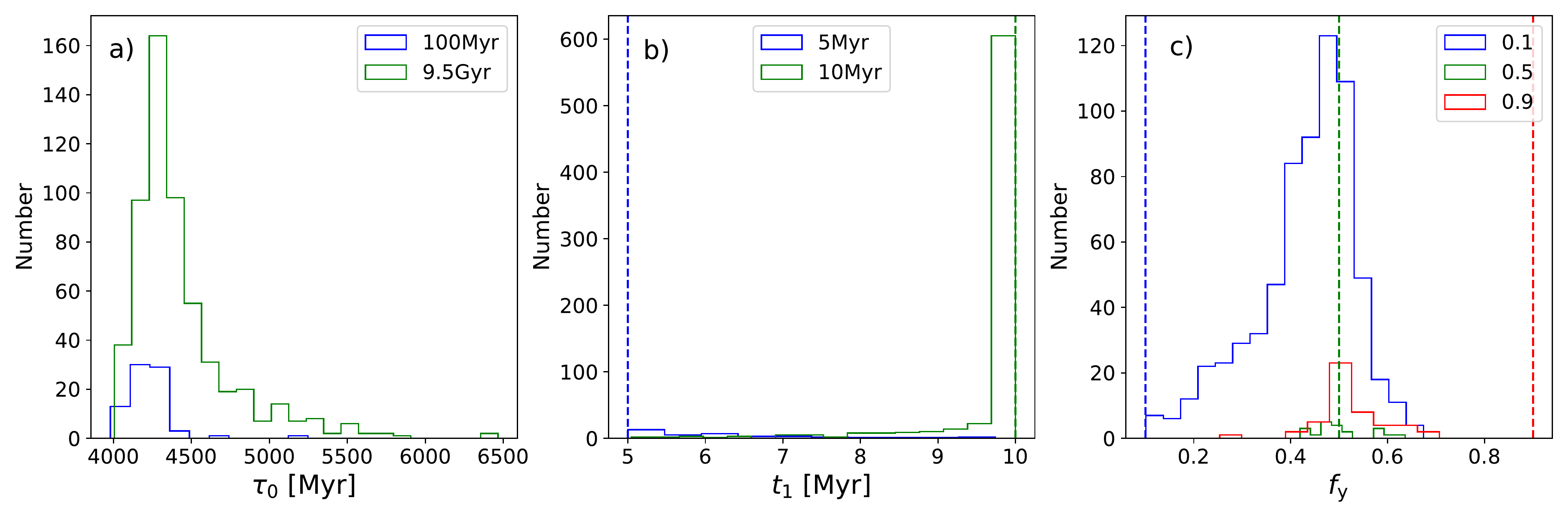}
    \caption{Histograms of the estimated parameters of 2exp SFH using mock catalogues. As 88\% of the sample has $\tau_\mathrm{0}$ best-fitted by 9.5 Gyr or 100 Myr, we chose to exclude other values from the plot. Parameters $t_\mathrm{0}$ and $\tau_\mathrm{1}$ are fixed. Each color represent a true value and its respective output distribution. To guide the eyes, we illustrate the position of each true value by dashed lines.}
    \label{fig:mock_2exp}
\end{figure*}

Figure \ref{fig:mock_2exp} shows the output parameters from 2exp SFH. In this case, the duration of the formation of old star has a large range of inputs. However we choose to analyze only 9.5 Gyr and 100 Myr, as 88\% of the sample is best-fitted by either of these values. Figure \ref{fig:mock_2exp}a exhibits how poor is the estimate of such parameter, with most of the return values around 4.3 Gyr, instead of 100 Myr or 9.5 Gyr. In Figure \ref{fig:mock_2exp}b, we verify that stars of 10 Myr have fairly well derived ages. Figure \ref{fig:mock_2exp}c presents the mass fractions of the late stellar population $f_\mathrm{y}$, we find a bad agreement between true and derived values. Most of objects have $f_\mathrm{y}\sim 0.5$, even if they should be 0.1 or 0.9. We concluded  that we cannot trust the mass fraction of young stars derived by 2exp SFH.

\begin{figure*}
\centering
\includegraphics[width=\textwidth]{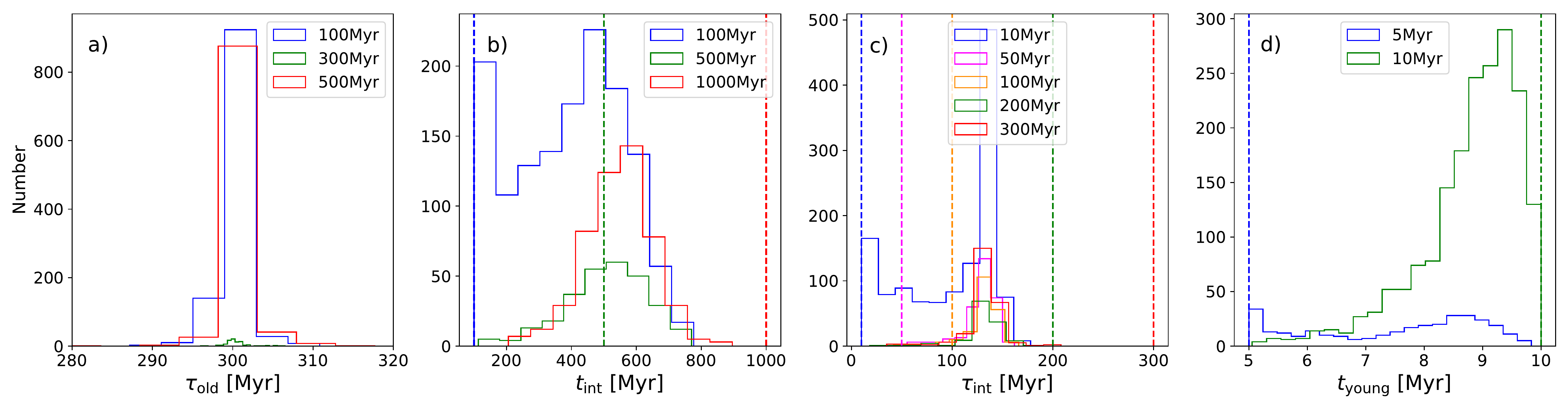}
\caption{Histograms of the estimated parameters of 3exp+mf SFH using mock catalogues. Parameters $t_\mathrm{old}$ and $\tau_\mathrm{young}$ are fixed. Each color represent a true value and its respective output distribution. To guide the eyes, we illustrate the position of each true value by dashed lines, except for $\tau_\mathrm{old}$ where the distributions are thin and centered at 300 Myr.}
\label{fig:mock_3exp+mf}
\end{figure*}

In Figure \ref{fig:mock_3exp+mf}, we analyze the output parameters from 3exp+mf SFH. Figure \ref{fig:mock_3exp+mf}a shows a poor reproduction of the duration of the oldest episode of star formation, with a thin distribution around 300 Myr for inputs of 100 Myr and 500 Myr. Figure \ref{fig:mock_3exp+mf}b demonstrates that stellar populations of 500 Myr have reliable measurements of ages, while the ones with 1 Gyr have underestimated values. Stellar populations of 100 Myr have similar probabilities of precise estimates and overestimated ages at 500 Myr. In Figure \ref{fig:mock_3exp+mf}c, we find that duration of star formation are overestimated for smaller inputs, 10 Myr, 50 Myr and 100 Myr, and underestimated for 200 Myr and 300 Myr. In Figure \ref{fig:mock_3exp+mf}d, the ages of the ionizing stellar population are poorly constrained, with an offset on the peak of this age distribution from its input of 10 Myr, and a secondary peak at 8.7 Myr for inputs of 5 Myr. The 3exp+mf SFH have two more fitted parameters not shown here, mass fraction of young and intermediate age stars. It is not a surprise that both of them have poor outputs. This SFH description associated to our set of broadband data does not provide a trustworthy fit solution.

\begin{figure*}
\centering
\includegraphics[width=0.9\textwidth]{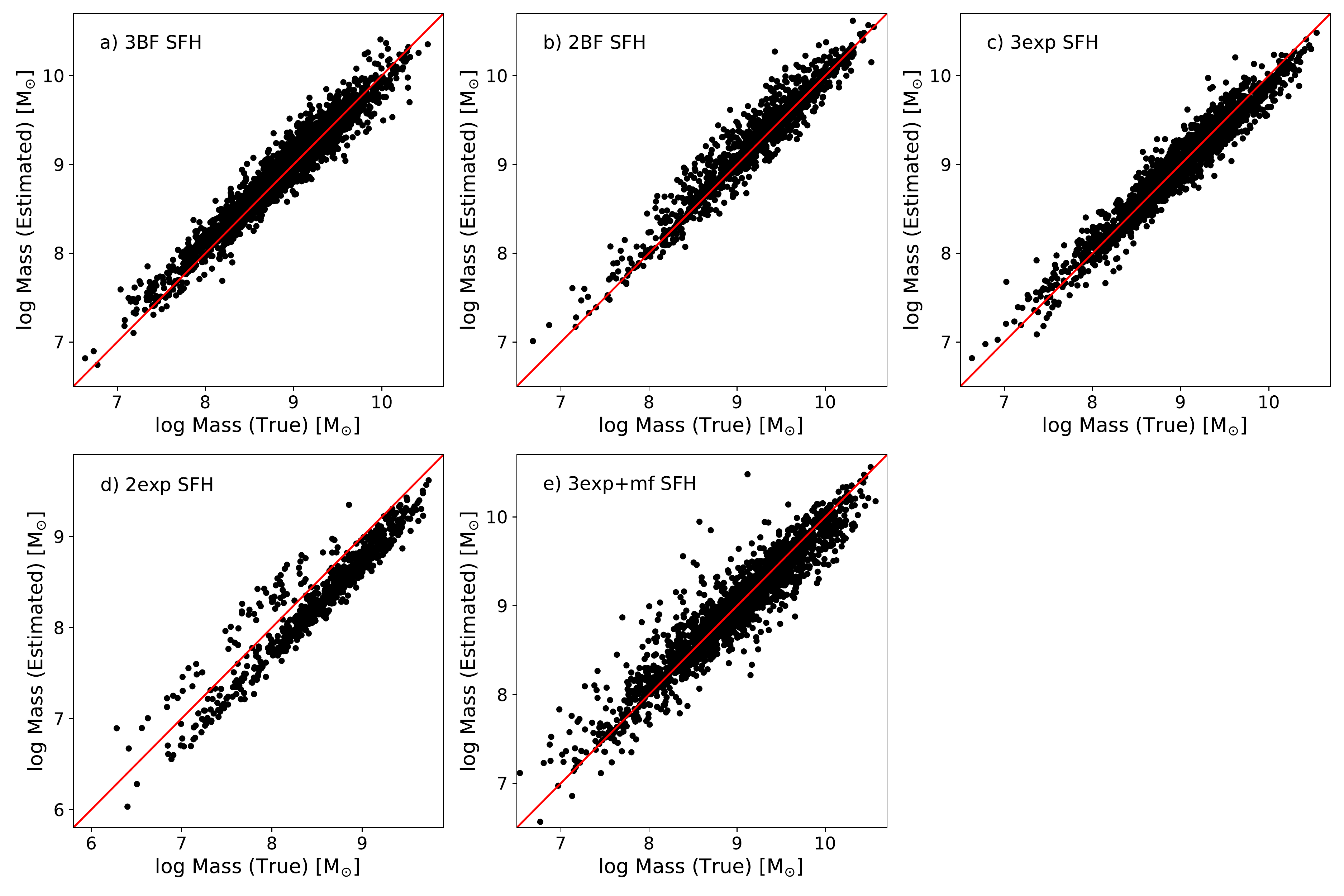}
\caption{\textbf{Estimated total stellar mass versus true mass from mock catalogues. Each panel presents the results from a given test. Upper panels from left to right: 3BF, 2BF and 3exp SFHs; lower panels: 2exp and 3exp+mf SFHs. The solid red lines represent the 1:1 relation to assist the plot interpretation.}}
\label{fig:mock_mass}
\end{figure*}

We proceed to check how well we can derive total stellar masses. Each panel in Figure \ref{fig:mock_mass} compares input (true) stellar masses with the estimate ones based on a given SFH. Figures \ref{fig:mock_mass}a, \ref{fig:mock_mass}b, \ref{fig:mock_mass}c and \ref{fig:mock_mass}e show that the derived masses present a good agreement with their true counterparts in 3BF, 2BF, 3exp and 3exp+mf SFHs, respectively. These panels present a similar scatter of $\sim$0.3 dex around the linear relation. Such spread agrees with the typical $1\sigma$ error associated to the estimated masses in this four SFHs. In Figure \ref{fig:mock_mass}d, we find that masses obtained by 2exp SFH are mostly underestimated from their inputs. Therefore, in general stellar masses have robust measurements for different SFHs, the exception is 2exp SFH.

All of these mock results indicate that the derived parameters are more robust for 3BF, 2BF and 3exp SFH than for 2exp and 3exp+mf SFHs. Thus, we decided to exclude 2exp and 3exp+mf SFHs from further analysis.

The 2BF SFH does not create intermediate age stars because it seems to favor intervals of star formation of < 1 Gyr, starting at 10 Gyr. Therefore, it does not reproduce the constraint (iii) set at the beginning of section 4. But due to the good fit results we will keep its result for general discussions.

Nevertheless, we find that the best description for HII galaxies are 3BF and 3exp SFHs, as they provide good fitting results and respect all constraints imposed by other independent measurements.

\begin{figure*}
\centering
\includegraphics[width=0.75\textwidth]{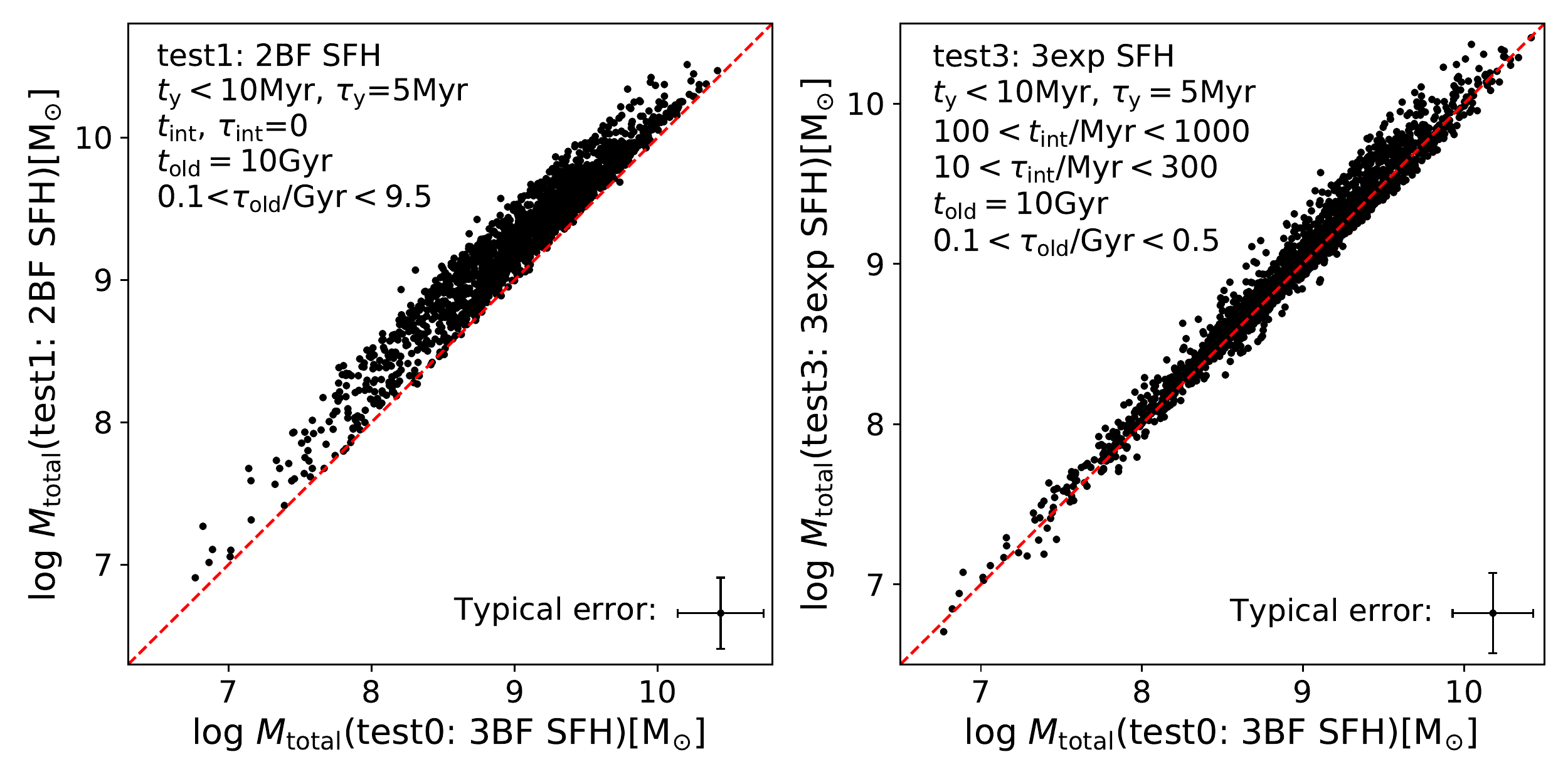}
\caption{Total stellar masses derived from: \texttt{test1} and \texttt{test3} compared to the ones from \texttt{test0}. \texttt{Test0} assumes a 3BF SFH with the following inputs: $t_\mathrm{y}<10$ Myr, $\tau_\mathrm{y}=5$ Myr, $100<t_\mathrm{int}\mathrm{Myr}<1000$, $10<\tau_\mathrm{int}/\mathrm{Myr}<300$, $t_\mathrm{old}=10$ Gyr and $0.1<\tau_\mathrm{old}/\mathrm{Gyr}<0.5$. The red dashed line is the 1:1 relation in both panels. Typical errors of stellar mass are $\sim 0.3$ dex in these three tests.}
\label{fig:total.masses}
\end{figure*}

\begin{figure*}
\centering
\includegraphics[width=0.75\textwidth]{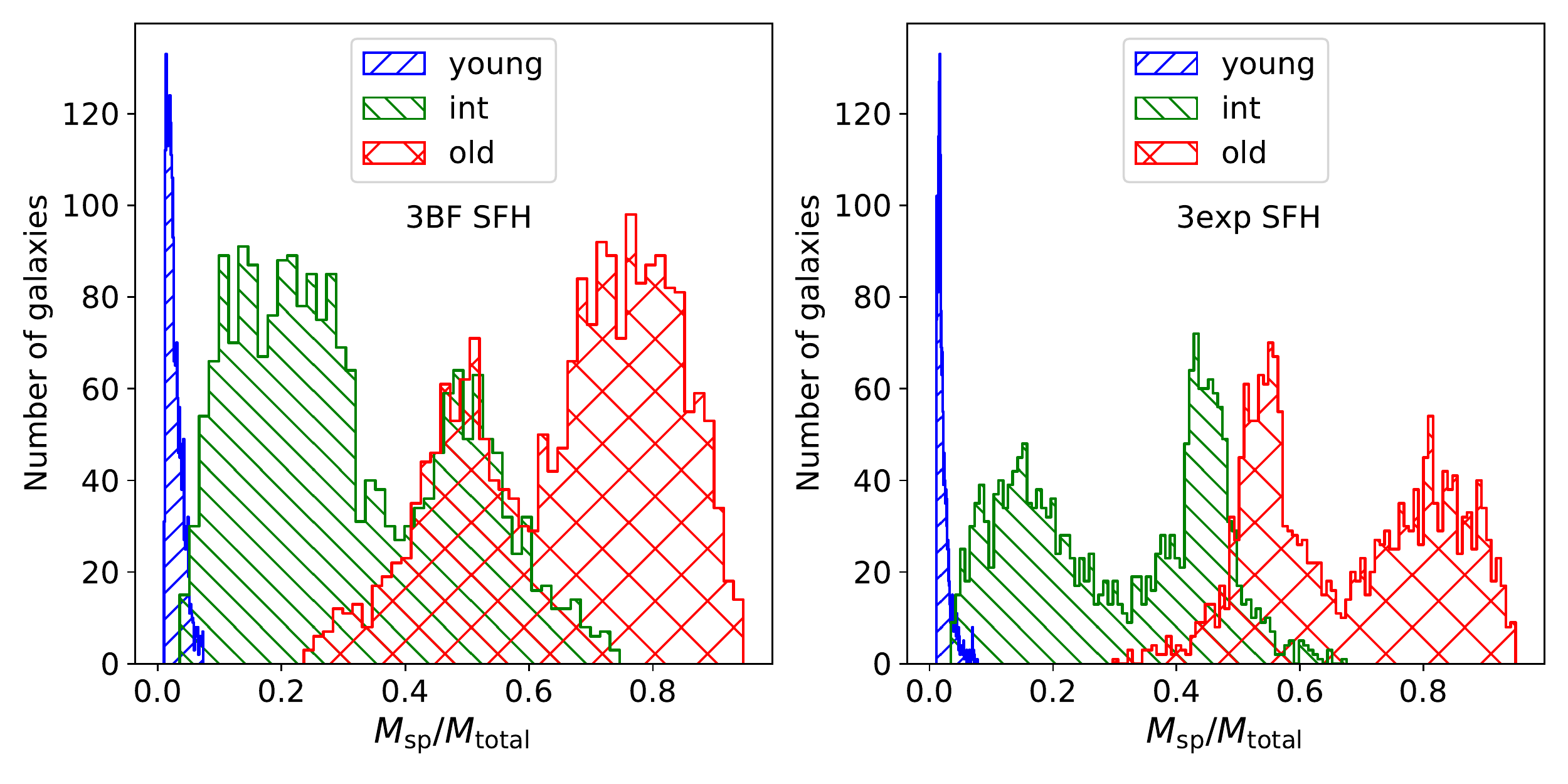}
\caption{Mass fraction for each stellar population (old: red, intermediate: green, young: blue) based on the SED fitting results from 3BF and 3exp SFHs. Both cases assume the same range of input parameters, see \texttt{test0} and \texttt{test3} in Table \ref{tab:sfh_input}.}
\label{fig:mass.fraction}
\end{figure*}

\subsection{Stellar mass results}
\label{subsec.mass}
By definition the total stellar mass is given by 
\begin{equation}
    M_\mathrm{total} = \int_{0}^{t_{0}} \mathrm{SFH}(t)\mathrm{\ }dt = t_{0} <\mathrm{SFR}>,
\label{eq.mass-def}
\end{equation}
where $t_{0}$ is the time when the star formation begins. Consequently different assumptions in the SFH result in the creation of different amounts of stars. Figure \ref{fig:total.masses} shows a comparison of the $M_\mathrm{total}$ from \texttt{test1} and \texttt{test3} in respect to \texttt{test0}. 

The left panel compares the results based on 2BF SFH. By excluding the intermediate age population, and allowing the star formation episode of the old population to extended up to 9.5 Gyr, the masses are shifted to higher values than 3BF SFH. Total masses from 2BF can be up to 0.5 dex more massive than 3BF. The difference in the stellar masses are directly related to how long the old population keeps creating stars. Most galaxies have a variation in $\log(M_\mathrm{total})$ of $\lesssim 0.3$ dex because the extension of star formation is < 1Gyr. Indeed, only $15\%$ of objects were fitted by very extended episodes of star formation, i.e > 1 Gyr, corresponding to differences in $\log(M_\mathrm{total})$ of 0.5 dex.

The right panel reveals the similarity on the $M_\mathrm{total}$ generated by 3BF and 3exp SFHs, described in Eqs. (\ref{eq:melnick}) and (\ref{eq:triple.exp}). The scatter in the relation between the derived mass from both tests is $\lesssim$0.3 dex that is less than the intrinsic median error of the masses, typically $\sim$0.3 dex. We concluded that a change in the shape of the SFH function does not strongly affect $M_\mathrm{total}$.

\begin{figure}
    \centering
    \includegraphics[width=0.47\textwidth]{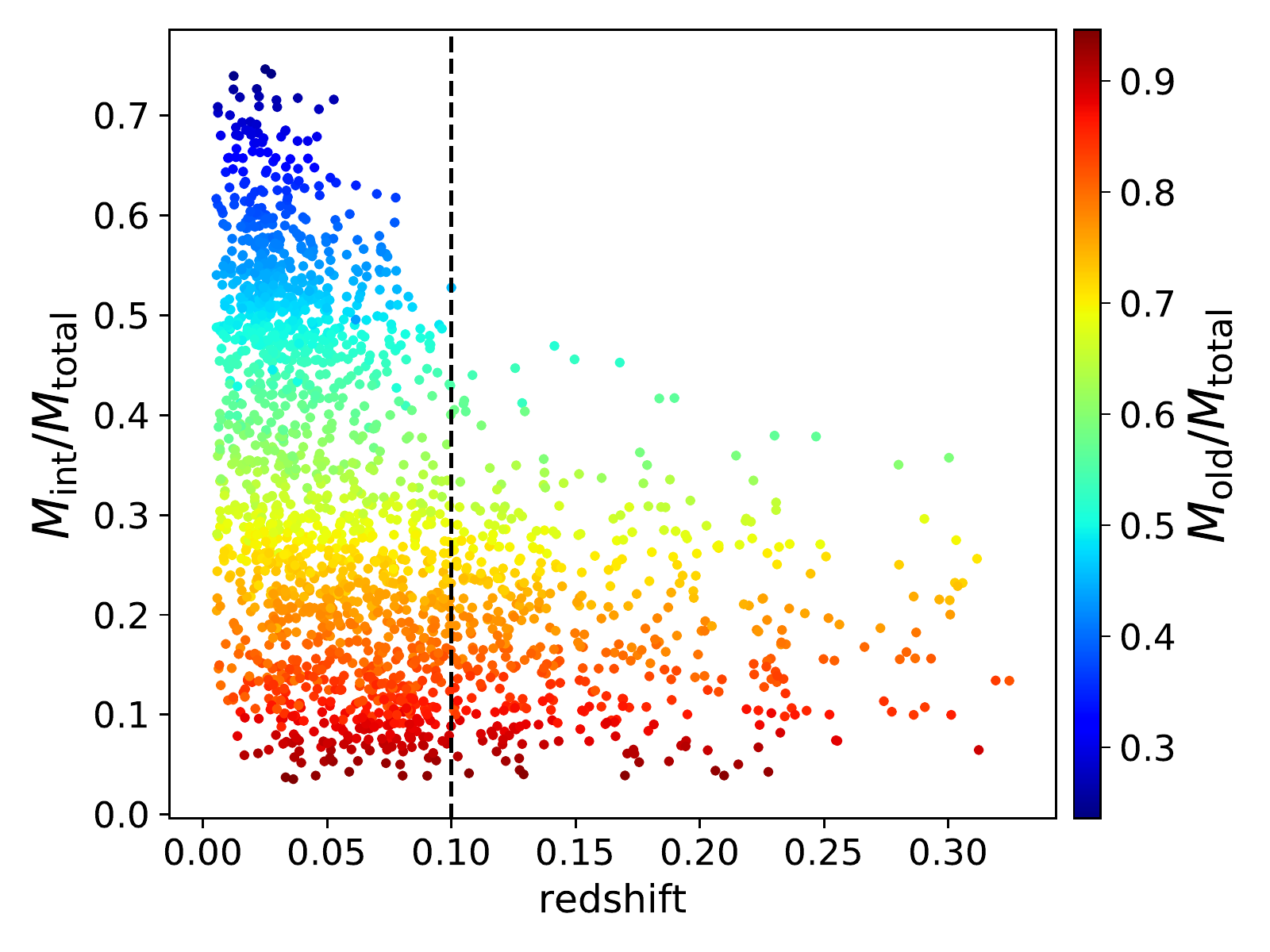}
    \caption{Mass fraction of intermediate age population from 3BF SFH versus redshift. The colorbar represents the mass fraction of old stars. The black dashed line signalizes redshift of 0.1, when intermediate age stars stop contributing with more than 50\% of total mass.}
   \label{fig:massfrac.z}
\end{figure}

\subsection{Stellar population results}
\label{subsec.stellar_pop}

CIGALE only outputs stellar masses for stars older and younger than a given input age separation. This separation can be set independent of the SFH assumed. Therefore, we we do not have a direct measure of the mass of intermediate population in a SFH with three star formation episodes. 

In order to estimate each stellar population mass we made two runs with the same SFH parameters but with different age separations: 10 Myr and 3 Gyr. The first run separates the young population mass, while the second one segregates the old stellar masses. We obtain the intermediate age stellar masses by subtracting the young and old stellar population masses from the total.

As 2BF SFH assumes two stellar populations, we only need one run with an age separation of 10 Myr. We find that young stars represent less than 10\% of the total amount of stellar mass, and by consequence the old stars dominate the total stellar mass.

Figure \ref{fig:mass.fraction} presents the ratio between the mass of each population in respect to the $M_\mathrm{total}$ from 3BF and 3exp SFHs. The left panel shows the results for 3BF SFH, in which young stars represent a small fraction of total mass, less than 10\%, while the contribution of intermediate age and old stars are more complex. Clearly, there are galaxies dominated by old stars, with 80\% or more of their mass. Nevertheless, there are about 20\% of galaxies with their total mass constituted by 50\% or more of intermediate stars.

The right panel in Figure \ref{fig:mass.fraction} presents the mass fraction results from \texttt{test3} where we alter the SFH description from 3BF to 3exp but kept the same input parameters. For the young stellar population the mass fraction remains small, <10\%. Although the total masses are not strongly affect by a change of shape, as seen in Figure \ref{fig:total.masses}, the contributions of each stellar population to these masses are modified. We find that the number of galaxies dominated by intermediate age stars (>50\% of $M_\mathrm{int}/M_\mathrm{total}$) decrease from 20\% (found in 3BF) to 6\% of the whole sample.

Complementary to the left panel in Figure \ref{fig:mass.fraction}, Figure \ref{fig:massfrac.z} demonstrates that galaxies with higher mass fraction of intermediate age stars, and consequently lower fraction of old population are seen only at lower redshifts, < 0.1. This result is for 3BF SFH, but a similar behavior is found for 3exp SFH with a smaller number of objects, as previously mentioned.

\begin{figure*}
    \centering
    \includegraphics[width=\textwidth]{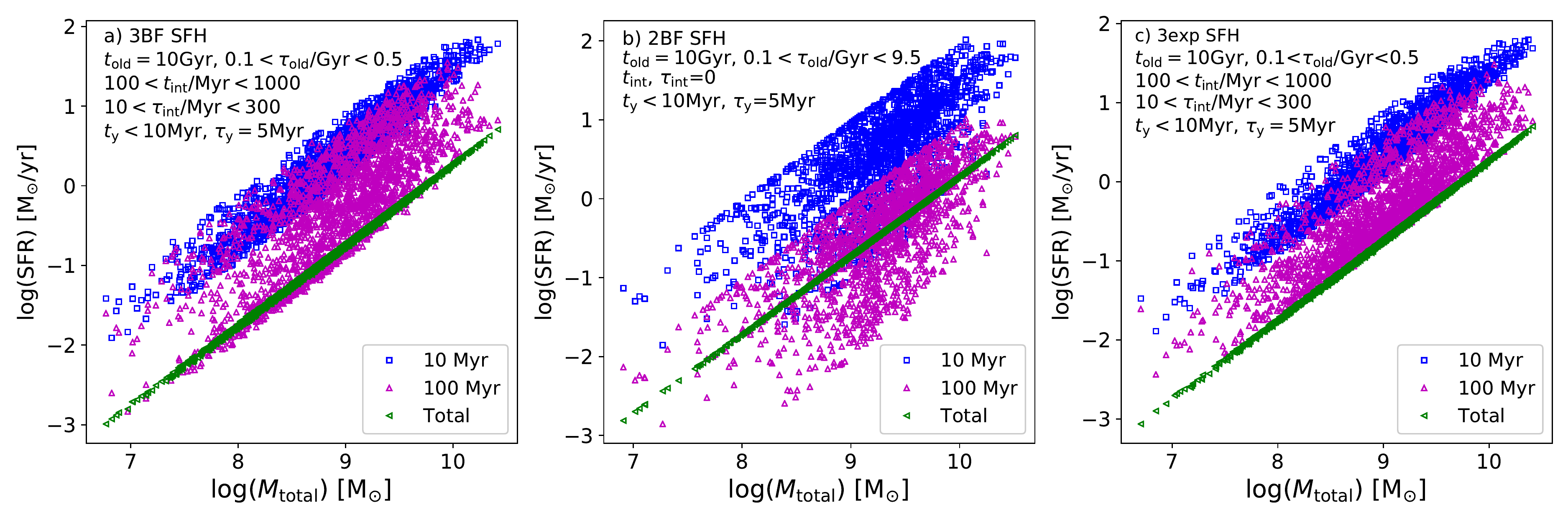}
    \caption{Star formation rate defined over an average of 10 Myr (blue squares), 100 Myr (magenta triangles) and the total SFH (green left triangles) versus total stellar mass for 3BF (left panel), 2BF (middle panel) and 3exp (right panel). The uncertainties in the measurements are typically of 30\% for both SFR and mass.}
    \label{fig:sfr_def}
\end{figure*}

All of these results reveal that the intermediate age stars can be a very important and significant component of HII galaxies, particularly at z<0.1.

\section{Main sequence of HII galaxies}
Star-forming galaxies define a narrow relation between their total masses and star formation rates generally known as Main Sequence (MS). In this section, we analyze the implications of the SFH in the MS for HII galaxies.

The Star Formation Rate (SFR) is defined as the mass of stars formed over a given time interval, such interval is not always explicitly informed in the literature, which can misguide the interpretation of such quantity. Different indicators are commonly used to infer the SFR such as H$\alpha$ luminosity, total infrared luminosity or ultraviolet luminosity among others. Each observable probes different time-scales that can vary from tens to hundred Myr, most of them being sensitive to scales smaller than 200 Myr \citep[]{2012ARA&A..50..531K}. The conversion of these observational indicators to SFR estimates are related to a number of assumptions such as spectral synthesis models, IMF, and SFH.

\citet{2019MNRAS.483.3213P} used a local SDSS spectroscopic galaxy sample to investigate the biases introduced by different SFR indicators and methods to locate the MS, and found that the slope of the relation can be strongly affected by the indicator used. \citet[]{2018A&A...616A.110E} showed that starbursts at $1.5<z<2.0$ have their SFR underestimated by SED-fitting when compared to the SFR obtained from UV plus IR light, with the dust attenuation being the main reason for such difference. Here, we only consider SFRs derived from SED modeling and fitting. By deriving the SFR and $M_\mathrm{total}$ from the same methodology, we avoid inconsistencies between these quantities due to different methods, adopting different assumptions. Such inconsistencies may lead to misinterpretation on the evolutionary paths of different galaxy types. 

We take advantage of CIGALE allowing estimates of SFR based on different intervals, to discuss the connection between the averaged time-scales and the SFHs in the SFR$-M_{*}$\footnote{Hereafter we will used $M_\mathrm{total}$ in place of $M_{*}$ to emphasize that the MS is always relative to the total mass over the SFH.} diagram for HII galaxies. 

Figure \ref{fig:sfr_def} shows the SFR$-M_\mathrm{total}$ relation for 3BF, 2BF and 3exp SFHs averaged on different time-scales: 10 Myr, 100 Myr, and total SFH. For SFR$_\mathrm{10 Myr}$, we are only estimating the strength of the recent burst, which according to Table \ref{tab:sfh_input} has to start between 5 and 10 Myr and to last 5 Myr. Furthermore, one has to bear in mind that a correlation between total SFR and $M_\mathrm{total}$ is always expected, since $M_\mathrm{total}$ is the time integral of star formation history (SFR(t)).

In Figures \ref{fig:sfr_def}a, we find that the SFR varies with the assumed averaged time-scale, for 3BF SFH. As expected, SFRs based on the last 10 Myr are higher than the others, due to typical young, ionizing stars in HII galaxies. Such SFRs correlates to total masses with a spread of less than 0.5 dex. SFRs averaged over 100 Myr present a drop in the amplitude and a larger scatter in comparison to SFR$_\mathrm{10Myr}$ as a consequence of no (or fewer) stars being formed between 10 Myr and 100 Myr. The total SFRs provide lower values and a tight relation with $M_\mathrm{total}$, which is explained by the imposition of all galaxies to begin star formation at the same time, i.e. 10 Gyr.

Figure \ref{fig:sfr_def}b presents how the different definitions of SFR behave in respect to the total mass, assuming 2BF SFH. As seen in Figure \ref{fig:sfr_def}a, the SFRs derived within 10 Myr are higher than the other defined SFRs. However, the spread found here, > 2 dex, is much larger than the one found in 3BF. SFRs averaged over the last 100 Myr are lower, in some cases even lower than the total SFR. This is a consequence of this particular SFH, where by definition there is no star formation between 10 and 100 Myr. The relation found between total SFR and mass is very thin.

\begin{figure*}
    \centering
    \includegraphics[width=\textwidth]{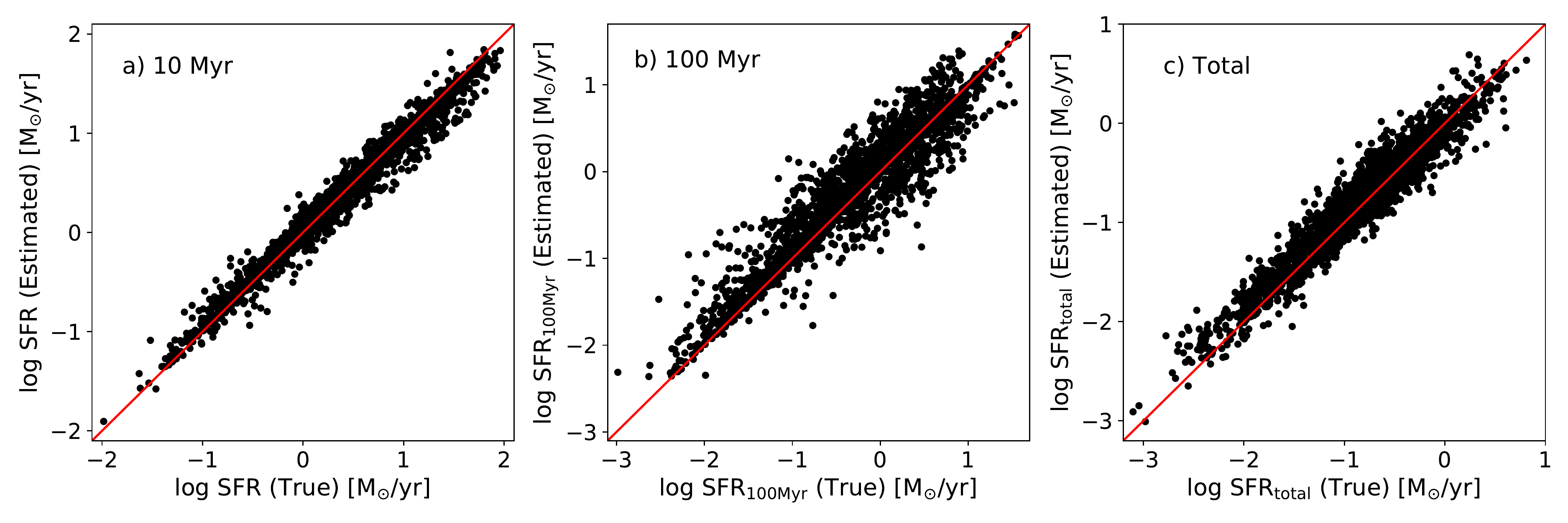}
    \caption{Estimated SFR versus true SFR for 3BF SFH, based on mock catalogues. Each panel presents results from a different different averaged time-scale: 10 Myr (left), 100 Myr (middle) and the total SFH (right). The red solid line represent the 1:1 relation.}
    \label{fig:mock_sfr}
\end{figure*}

\begin{figure*}
\centering
\includegraphics[width=0.75\textwidth]{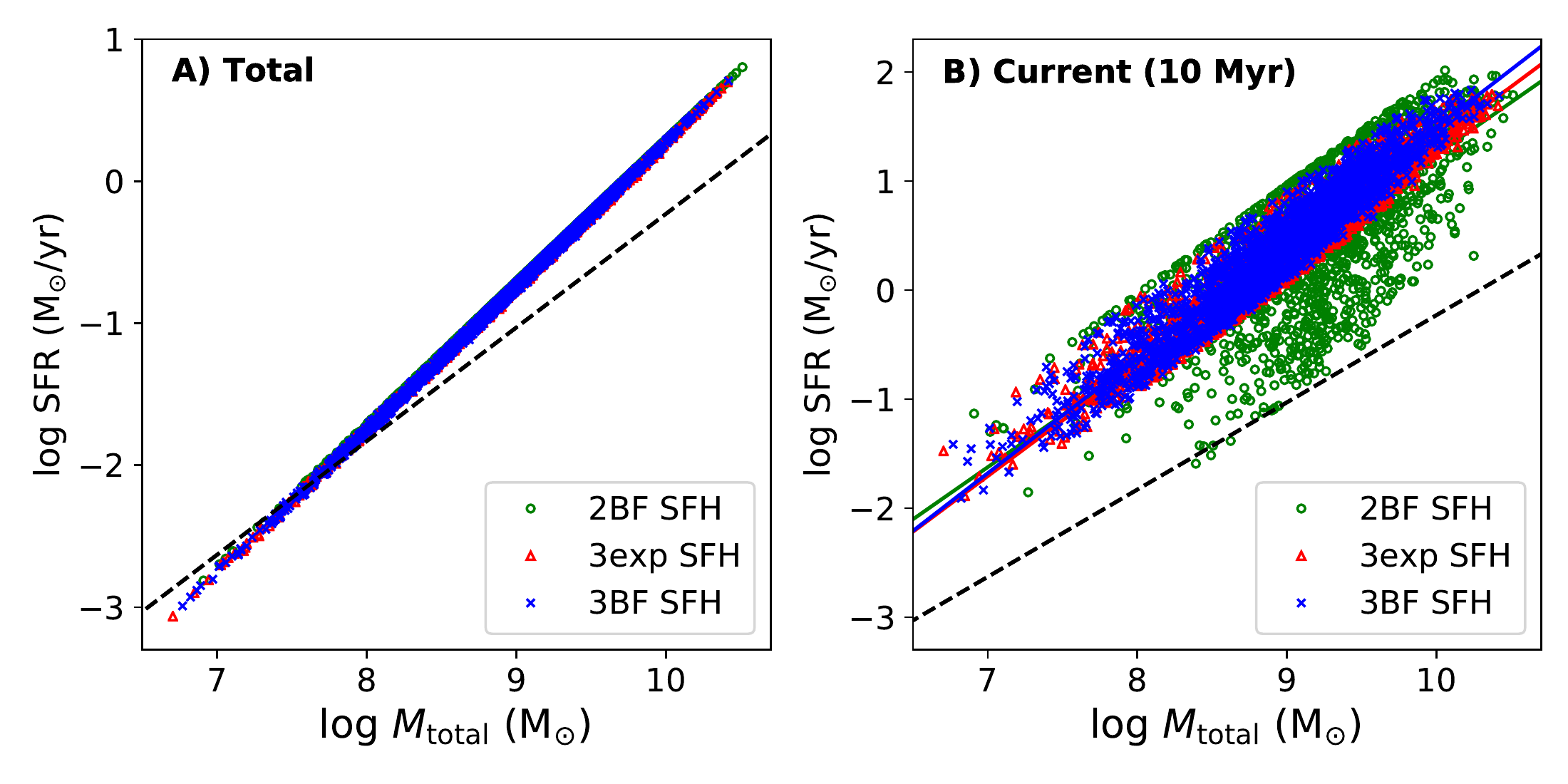}
\caption{Panel A: Total SFR versus stellar mass from results based on different SFH assumptions: 3BF (blue crosses), 2BF (green circles) and 3exp SFHs (red triangles). Panel B: Current SFR versus total mass for each SFH.  The typical errors for the log($M_\mathrm{total}$) and log(SFR) are both $\sim$0.3 dex, for all assumed SFH. The best-fit of each SFH is represented by a solid line of its respective color, as in legend. The black dashed line is based on \citet{2004MNRAS.351.1151B} data.}
\label{fig:main_seq_comp}
\end{figure*}

Figure \ref{fig:sfr_def}c illustrates the same three definitions of SFR, but based on 3exp SFH. Similar to the results from 3BF SFH seen in Figure \ref{fig:sfr_def}a, time-scales of 10 Myr produce higher SFRs with smaller spread, while the ones of 100 Myr yield lower SFRs with larger spread. Total SFRs provide the lowest values of the three definitions, and a very tight relation with mass.

In order to give credibility to our SFR discussion, Figure \ref{fig:mock_sfr} compares outputs and inputs of different definitions of SFR, assuming 3BF SFH. These results were obtained following the procedure introduced in Subs. \ref{subsec.mock_analysis} based on mock catalogues. We chose to discuss results from 3BF SFH, but similar behaviors are found for 3exp SFH, and even better ones for 2BF.

In Figure \ref{fig:mock_sfr}a, we see a good correspondence between true and derived SFRs averaged over 10 Myr. Meanwhile, Figure \ref{fig:mock_sfr}b shows a larger spread around the 1:1 relation for 100 Myr, with most values overestimated in respect to the inputs. Figure \ref{fig:mock_sfr}c demonstrates a fairly good correlation between true and obtained total SFRs, where the spread is less than the uncertainties of the SFR, i.e. 0.3 dex.

\subsection{Total and current SFR for HII galaxies}
In order to understand the effects of SFH in the SFR$-M_\mathrm{total}$ diagram we need to compare results from different SFH descriptions. Panel A in Figure \ref{fig:main_seq_comp} compiles SFR$_\mathrm{total}$ based on the results from: 3BF (\texttt{test0}); 2BF (\texttt{test1}); and 3exp (\texttt{test3}).

From the previous section we know that an observational constraint for HII galaxies is the age of the oldest star. By imposing age=10 Gyr for different SFHs we find that their SFR$-M_\mathrm{total}$ results are in close agreement, almost falling upon each other. The main difference is the total mass range which follows the relation presented in Figure \ref{fig:total.masses}.

The amount of mass created will dependent on the shape of the SFH and the duration of the episode(s), i.e. the integration time as shown in Eq.(\ref{eq.mass-def}). If we assume a similar star formation time, i.e. the SFR is averaged over the same interval, the galaxies will change their position following a similar path with a small variation associated to the different SFHs at a given mass. This interpretation agrees with \citet[]{2017ApJ...844...45O} who argued that if the bulk of galaxies are roughly coeval, then only the shape of SFHs would set the scatter in the relation, at fixed mass. 

In the cases where the star formation time is established as 10 Gyr, the MS becomes in close agreement with the results found by other works such as \citet{2004MNRAS.351.1151B} and \citet{2015ApJS..219....8C} but with different slopes. This proximity with the MS from other datasets based on ``normal" star forming galaxies was unexpected, because HII galaxies are starburst systems, characterized by high SFR for their masses. Studies based on starburst-type galaxies suggest that these objects typically lie above the MS \citep[e.g.][]{2011A&A...533A.119E,2011ApJ...739L..40R,2012ApJ...747L..31S}. Although \citet[]{2018A&A...616A.110E} discussed the possibility of a number of starburst galaxies being hidden in the MS due to the compactness of their starburst regions. 

As the SFH of HII galaxies expands along 10 Gyr with a complex mixture of populations, the total SFR becomes similar to ``normal" star forming galaxies. \citet{2017ApJ...851...22M} point out that the current SFR may fluctuate, but if the SFR$_\mathrm{current}$ >  mean SFR extends for long, it increases SFR$_\mathrm{total}$ but also builds up stellar mass. In other words, a very high SFR, if sustained, makes enough stars to drive back the galaxy to a MS of slope close to unity. Of course, there can be temporary excursions with SFR$_\mathrm{current}$>SFR$_\mathrm{total}$. 

The behavior of the MS is still debatable in the literature. If we assume a single power law the values of the slope vary widely with the SFR indicator and/or with the selected sample \citep[for a detailed discussion see][]{2014ApJS..214...15S}. However, several authors suggest that the SFR$-M_\mathrm{total}$ relation is not a single power law but it exhibits a curvature or a broken power law with a flatter slope at the high-mass relative to the low-mass regime \citep[][]{2015ApJ...801...80L,2015A&A...575A..74S,2015ApJ...811L..12W,2016ApJ...817..118T,2019MNRAS.483.3213P}. Indeed, \citet[][]{2015ApJ...801L..29R} discussed that the pre-selection of a sample could partially be the cause of the bending at high masses. 

In our analysis restricted to HII galaxies at $z<0.4$, we find a single power-law with slopes close to unity for all SFH in Panel A from Figure \ref{fig:main_seq_comp}. Indeed, we expected slopes of unity because the MS compares mean SFH (total SFR) with an integral of the SFH (total mass), both estimated by SED fitting. Therefore, we conclude that the slope of this relation does not provide any information about the SFH.

One should always be careful to apply the same assumptions to estimate the mass and the SFR. Any calibration of SFR has implicitly choices of IMF, stellar synthesis models, SFH (or at least an age for simple stellar populations). If the assumptions used to obtain the SFR differ from the ones used to estimate the stellar mass, the slope of the MS could be affected and be different than one.

In order to be sensitive to the high current SFR, we should consider the SFR average over the last 10 Myr. In Panel B from Figure \ref{fig:main_seq_comp} we represent the SFR$_\mathrm{10Myr}$ for different SFHs. Although the SFR$_\mathrm{total}$ follows a similar path for different SFHs assuming the same age, the current SFR exhibits the expected starburst behavior with HII galaxies being placed above the MS for most histories applied. The zero point and spread in this relation is a direct consequence of the SFH, reflecting the fraction of young stars in respect to the total amount of stars in a galaxy. 

The differences found between panel A and B in Figure \ref{fig:main_seq_comp} is a consequence of SFR$_\mathrm{10Myr}$ being greater than the mean SFR over the whole SFR (10 Gyr).

\begin{figure}
\centering
\includegraphics[width=0.47\textwidth]{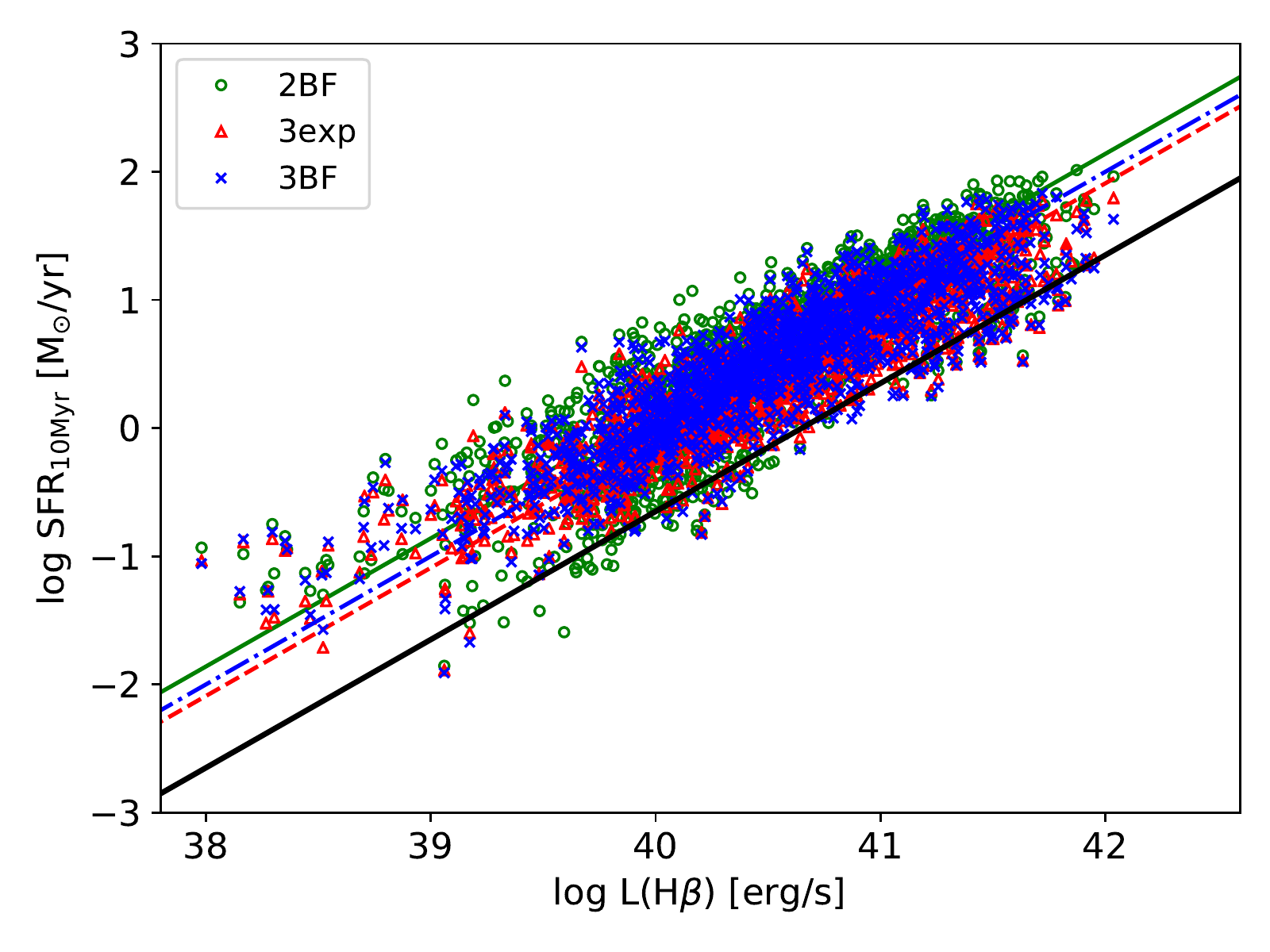}
\caption{Current SFR versus the H$\beta$ luminosity. The lines represent the best-fit of each SFH result. The solid thick black line is the relation by \citet[]{1998ARA&A..36..189K} based on normal spiral galaxies. The typical errors for log(SFR$_\mathrm{10Myr}$) and log L(H$\beta$) are $\sim$30\% and $\lesssim$10\%, respectively. Symbols as in legend.}
\label{fig:lum_sf}
\end{figure}

In order to infer the efficiency in forming of stars in recent times, we estimate the specific SFR, sSFR=SFR/$M_\mathrm{*}$ only in the last 10 Myr. Following \citetalias{2018A&A...615A..55T}, we calibrate the current SFR in respect to the observed $L(\mathrm{H}\beta)$ for 2BF, 3exp and 3BF, resulting in
\begin{equation}
\log(\mathrm{SFR})=-39.97\pm0.29 + \log L(\mathrm{H}\beta) \, (rms=0.25),
\label{eq.sfr2bf_cal}    
\end{equation}
\begin{equation}
\log(\mathrm{SFR})=-40.10\pm0.21 + \log L(\mathrm{H}\beta) \, (rms=0.20),
\label{eq.sfr3exp_cal}    
\end{equation}
\begin{equation}
\log(\mathrm{SFR})=-40.02\pm0.21 + \log L(\mathrm{H}\beta) \, (rms=0.22),
\label{eq.sfr3bf_cal}    
\end{equation}
respectively. Figure \ref{fig:lum_sf} shows the SFR$_\mathrm{10Myr}$ versus $L(\mathrm{H}\beta)$ for different SFHs, and their best-fit. Also we include the commonly used \citet[]{1998ARA&A..36..189K} calibration for comparison. The differences in the zero point from our calibrations to Kennicutt relation is mostly due to the fact that we isolate the starburst component. 

\begin{figure}
    \centering
    \includegraphics[width=0.47\textwidth]{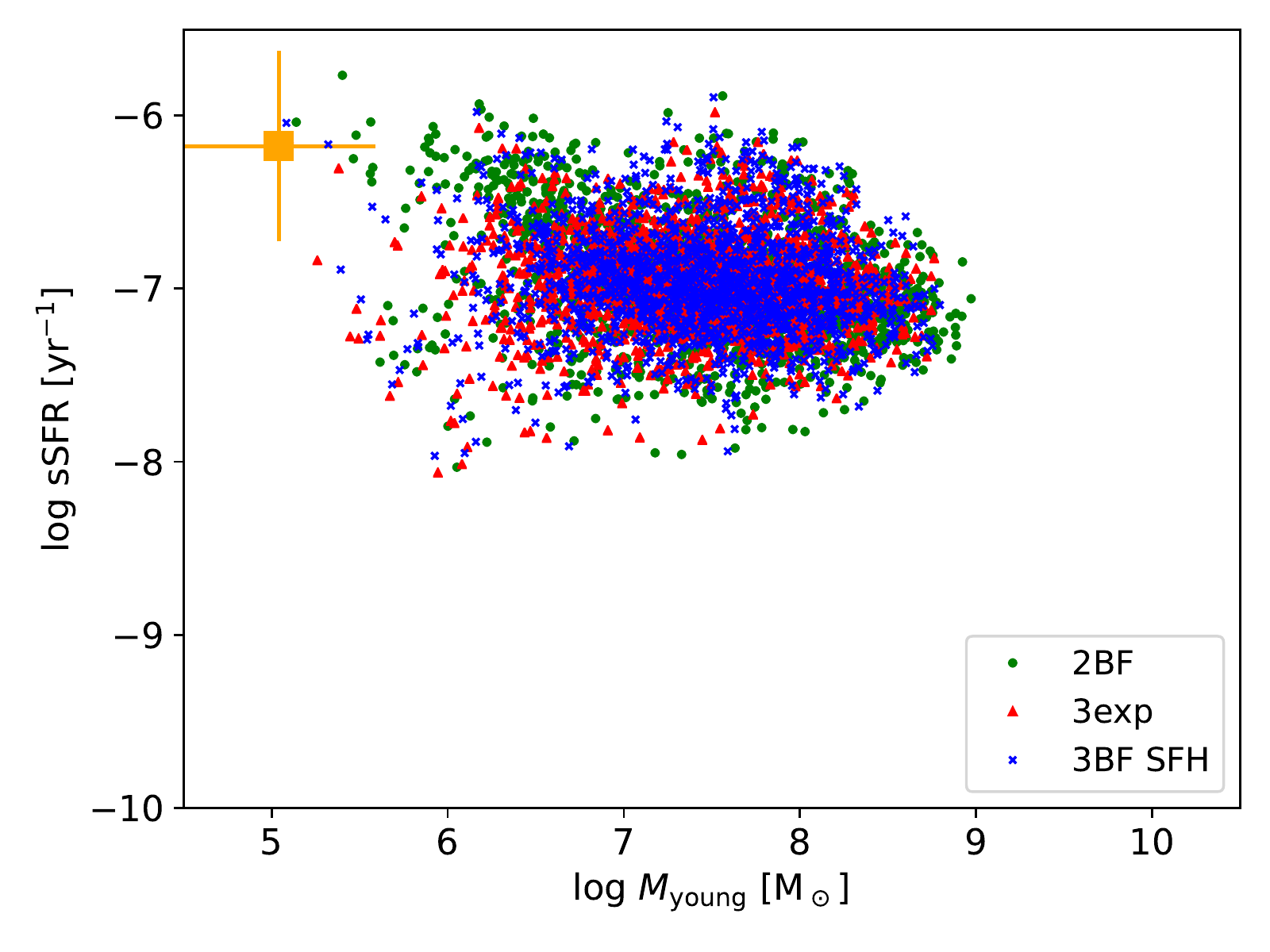}
    \caption{Specific SFR versus stellar mass for the recent burst assuming 2BF, 3exp and 3BF SFHs.  The typical errors for log(sSFR) and log($M_\mathrm{young}$) are $\sim$30\% in both cases. The orange square represents the Giant HII Region 30Dor in the LMCs \citep[][]{2013A&A...558A.134D}. }
    \label{fig:sSFR_y}
\end{figure}

Based on the fitted SFRs we can re-obtain the current SFR and divide by the the stellar mass created in the last 10 Myr. Figure \ref{fig:sSFR_y} shows the derived sSFR versus $M_\mathrm{young}$ for the three SFHs. These sSFR results demonstrate that the efficiency in the production of stars are consistent among different SFHs.

Figure \ref{fig:sSFR_y} also presents the comparison between the sSFR results with an independent measurement of star formation derived for the Giant HII Region 30 Doradus (hereafter, 30 Dor) in the Large Magellanic Clouds \citep[][]{2013A&A...558A.134D}, using a Lyman continuum census and a \citet[][]{2001MNRAS.322..231K} IMF, as also shown in \citetalias{2018A&A...615A..55T}. As the closest example of a simple stellar population with a star formation spread over 5 Myr \citep[]{1999A&A...347..532S}, 30 Dor is an upper limit to the efficiency of star formation in starbursts. Our results are consistent with 30 Dor independent measurement.

In particular, low mass starbursts in HII galaxies have sSFR reaching that of 30 Dor. Thus, HII galaxies are the simplest starbursts in galactic scale with the highest star formation efficiencies.

\section{Conclusions}
We performed a SED analysis assuming SFHs with different shapes and input parameters to search for the best analytical description of HII galaxies. Our sample contains local objects (z<0.4) selected by SDSS spectroscopy, and photometric data from GALEX, SDSS, UKIDSS, and WISE surveys. 

Combining our $\chi^{2}_\mathrm{red}$, BIC and mock analysis from the SED fitting we find that
\begin{itemize}
    \item double exponential function yields the worst fit of all tests, thus is excluded;
    \item SFHs containing three star formation episodes provide good fits to the observational data;    
    \item three episodic SFH with a varying amplitude does not give trustworthy outputs, then is eliminated;
    \item a constant extended SFR started at 10 Gyr, followed by a quiescent period with a recent burst does not provide bad fitting results, but lacks intermediate age stars. 
\end{itemize}
Therefore, we confirm the findings of \citetalias{2018A&A...615A..55T} that HII galaxies are better described by SFHs following exponentially declining or constant functions containing at least three star formation episodes: an old (10 Gyr); an intermediate (100$-$1000 Myr); and a young stellar population where the most recent episode must have ages < 10 Myr.

We analyzed the so-called main sequence of star formation (MS), the relation between SFR and total stellar mass ($M_\mathrm{total}$) for HII galaxies using the values derived from SED fitting. The main conclusions are
\begin{itemize}
   \item for a fixed age of 10 Gyr different SFHs produce only minor variations in the MS. HII galaxies will move along a similar path depending on the shape(s) or duration of the star formation episode(s);
  \item the slope of the SFR$_\mathrm{total}-M_\mathrm{total}$ relation does not provide any information about the SFH, with a close to unity value for all cases discussed here;
    \item HII galaxies only lie above the local MS when the SFR is averaged over the last 10 Myr.
\end{itemize}

The star forming main sequence can not be used as a tool to provide additional information about the SFH of HII galaxies. The excursions from the MS depend on the detailed description of the SFH, not only the average over Hubble time, and total mass and SFR must be jointly estimated consistently to avoid mis-interpretations on galaxy evolution. 

\section*{Acknowledgements}
A.R.L acknowledges the financial support from CNPq through the PCI fellowship. J.M. acknowledges support from CNPq Ciencia Sem Fronteiras grant at the Observatorio Nacional in Rio de Janeiro, and the hospitality of ON as a PVE visitor. Finally, we thank the anonymous referee for the comments and suggestions that improved the paper.

\section*{Data availability}
The data underlying this article will be shared on reasonable request to the corresponding author.











\bsp	
\label{lastpage}
\end{document}